\documentclass{emulateapj}

\slugcomment{Accepted version from August 25, 2018}
\shorttitle{Abundances in the LA towards NGC\,3783}
\shortauthors{P. Richter et al.}

\begin{document}

\title{
New constraints on the nature and origin of the Leading Arm\\
of the Magellanic Stream
}

\author{Philipp Richter\altaffilmark{1},
Andrew J. Fox\altaffilmark{2},
Bart P. Wakker\altaffilmark{3},
J. Christopher Howk\altaffilmark{4}
Nicolas Lehner\altaffilmark{4},
Kathleen A. Barger\altaffilmark{5},
Elena D'Onghia\altaffilmark{3,6},
\and
Felix J. Lockman\altaffilmark{7}
}

\affil{$^1$Institut f\"ur Physik und Astronomie, Universit\"at Potsdam,
Haus 28, Karl-Liebknecht-Str.\,24/25, 14476 Golm (Potsdam),
Germany}
\affil{$^2$Space Telescope Science Institute, Baltimore, MD 21218, USA}
\affil{$^3$Department of Astronomy, University of Wisconsin-Madison,
475 N. Charter Street, Madison, WI\,53706, USA}
\affil{$^4$Department of Physics, University of Notre Dame, 225 Nieuwland Science Hall,
Notre Dame, IN 46556, USA}
\affil{$^5$Department of Physics and Astronomy, Texas Christian University, TCU Box 298840, 
Fort Worth, TX 76129, USA}
\affil{$^6$Center for Computational Astrophysics, Flatiron Institute, 162 Fifth Avenue, 
New York, NY 10010, USA}
\affil{$^7$Green Bank Observatory, P.O. Box 2, Route 28/92, Green Bank, WV 24944, USA}

%

\begin{abstract}

We present a new precision measurement of gas-phase abundances of S, O, N, Si, Fe, P, Al, Ca
as well as molecular hydrogen (H$_2$) in the Leading Arm (region II, LA\,II) of the
Magellanic Stream (MS) towards the Seyfert galaxy NGC\,3783. The results are based on high-quality
archival ultraviolet/optical/radio data from various different instruments
(HST/STIS, FUSE, AAT, GBT, GB140\,ft, ATCA). Our study updates previous results from
lower-resolution data and provides for the first time a self-consistent
component model of the complex multi-phase absorber, delivering
important constraints on the nature and origin of LA\,II.
We derive a uniform, moderate $\alpha$ abundance in the two main absorber groups at 
$+245$ and $+190$ km\,s$^{-1}$ of $\alpha$/H$\,=0.30\pm 0.05$ solar, a low nitrogen abundance of
N/H$\,=0.05\pm 0.01$ solar, and a high dust content with substantial dust depletion values for 
Si, Fe, Al, and Ca.
These $\alpha$, N, and dust abundances in LA\,II are similar to those observed in the 
Small Magellanic Cloud (SMC). From the analysis of the H$_2$ absorption, we determine
a high thermal pressure of $P/k\approx 1680$ K\,cm$^{-3}$ in LA\,II, in line with the idea
that LA\,II is located in the inner Milky Way halo at a $z$-height of $<20$ kpc where it 
hydrodynamically interacts with the ambient hot coronal gas.
Our study supports a scenario, in which LA\,II stems from the break-up of a metal- and dust-enriched 
progenitor cloud that was recently ($200-500$ Myr ago) stripped from the SMC.

\end{abstract}

%

 
%

\section{Introduction}

The gravitational and hydrodynamical interaction between the Milky Way (MW) 
and the Magellanic Clouds (MCs) and the subsequent star-formation activity in the 
Clouds have transported more than one billion solar masses of gas from the MCs into 
the circumgalactic medium (CGM) of the Milky Way. 
These processes have produced extended gaseous streams and clouds at distances of
$d\approx 20-100$ kpc that cover more than $40$ percent of the sky (Wannier et al.\,1972;
Mathewson et al.\,1974; Putman et al.\,1998; Br\"uns et al.\,2005; Fox et al.\,2014;
Richter et al.\,2017). This gigantic circumgalactic gas reservoir dominates the Milky Way's current
and future gas accretion rate (see recent reviews by D'Onghia \& Fox 2016; Putman et al.\,2012; Richter 2017)
and thus has a major impact on the Galaxy's evolution. The neutral gas bodies 
of these CGM clouds and streams
can be observed in H\,{\sc i} 21\,cm emission, while the more diffuse and more extended 
ionized gaseous envelopes can be traced in ultraviolet (UV) metal absorption against distant,
extragalactic point sources (such as quasars and other type of AGN, for simplicity 
hereafter referred to as QSOs) as well as in H$\alpha$ emission (see, e.g., Richter 2017;
Barger et al.\,2013, 2017).

Absorption-line studies of the Milky Way gas environment, in particular, can be readily compared to 
similar studies of the CGM around other, more distant galaxies at
low and high redshift (e.g., Werk et al.\,2013; Stocke et al.\,2014; Liang \& Chen 2014;
Richter et al.\,2016).
Detailed studies of the spatial distribution and chemical composition of the Milky Way's 
CGM provide important clues on the on-going formation
and evolution of the Galaxy in its Local Group environment through gas infall and
satellite interaction. They also deliver detailed constraints on the filling factor 
and physical conditions of multi-phase circumgalactic gas around star-forming disk galaxies 
and thus are of high relevance to evaluate the importance of the CGM 
for galaxy evolution, in general. 

The most prominent of the circumgalactic gas features generated by the interaction
between the MCs and the MW is the Magellanic Stream (MS), 
an enormous stream of neutral and ionized gas in the southern hemisphere extending over 
more than 200 degrees (Nidever et al.\,2010), and possibly over several hundred
kpc in linear size (D'Onghia \& Fox 2016). The main body of the MS is believed to be at a distance of 
$50-100$ kpc, thus located in the outer Milky Way halo. The spatial extension of 
the MS into the northern sky at $l>300$ is
the so-called Leading Arm (LA), a conglomerate of scattered clouds seen
in 21\,cm data and in UV absorption whose ionized enveleope possibly extends far into the 
northern sky (Fox et al.\,2018; Richter et al.\,2017). 
It was previously argued that the LA must be of purely tidal origin, as it leads the 
orbital motion of the Magellanic System (which includes the MS, the LA, and the Magellanic
Bridge) around the MW (e.g., Putman et al.\,1998). However, more recent hydrodynamical
simulations indicate that only the combination of multiple tidal stripping events, outflows 
from the Magellanic Cloud, and ram-pressure stripping can explain the complex spatial
distribution of the MS and LA gas components and the observed abundance variations 
therein (Besla et al.\,2010, 2012; Fox et al.\,2018; Pardy et al.\,2018).
The LA, which is sub-divided into four main substructures, LA\,I-IV (Putman et al.\,1998; Br\"uns et
al.\,2005; Venzmer et al.\,2012; For et al.\,2012) is much closer than the MS ($d<20$ kpc), 
as evident from the analysis of young stars that recently have been found in the LA
(Casetti-Dinescu et al.\,2014; Zhang et al.\,2017). 

In our previous studies, we have used UV absorption-line data for many dozens 
of QSO sightlines together with 21\,cm emission-line data from various instruments to
characterize the overall chemical composition of the MS and the LA, to estimate their 
total mass, and to pinpoint their contribution to the Milky Way's gas-accretion rate 
(Fox et al.\,2010, 2013, 2014, 2018; Richter et al.\,2013, 2017). 
In addition to these large surveys, detailed analyses for individual sightlines are likewise 
highly desired. Such studies provide a tremendous amount of accurate information 
on the chemical enrichment pattern and the local physical conditions in the gas,
which is very important to discriminate between different scenarios for
the origin of the MS and LA (e.g., LMC vs.\,SMC).
For instance, the substantially higher metallicity and enhanced dust abundance found in the MS
along the Fairall\,9 sightline (Richter et al.\,2013; hereafter referred to as R13) 
compared to other MS sightlines (Fox et al.\,2013) demonstrates that 
the trailing arm of the MS has a dual origin with spatially and kinematically distinct
filaments stemming from both LMC {\it and} SMC (see also Nidever et al.\,2008, 2010). 
This result sets important constraints for dynamical models
of the LMC/SMC/MW gravitational interactions and provides independent evidence for 
an enhanced star-formation activity in the LMC that has lifted $\alpha$-enriched gas
into the Milky Way halo (see discussions in R13; Bustard et al.\,2018; Pardy et al.\,2018).
The stellar activity of the LMC is currently driving over $10^7$ solar masses out of this 
galaxy in a large-scale galactic wind (Barger 2016) and similar winds were likely generated 
during episodic periods of elevated star formation in the past.

Following our long-term strategy to explore the origin and fate of the Magellanic System
and its role for the past and future evolution of the Milky Way and the Local Group (LG),
we here present a detailed study of the chemical abundances and physical conditions in
the LA\,II along the line of sight towards NGC\,3783. Because of the brightness of the 
background Seyfert galaxy NGC\,3783 and the high neutral gas column density, 
the NGC\,3783 direction is the best-studied 
QSO sightline for absorption-line measurements in the LA with a number of
detailed studies (Lu et al.\,1994, 1998; Sembach et al.\,2001; Wakker et al.\,2002, 
hereafter WOP02; Wakker 2006)
using medium-resolution UV data from the early-generation HST/UV spectrograph
GHRS (Goddard High Resolution Spectrograph), data from the {\it Far Ultraviolet
Spectroscopic Explorer} (FUSE),
and 21\,cm inferometer data from the Australian Compact Telescope Array (ATCA).
The best available UV data set (in terms of spectral resolution and S/N) for NGC\,3783
was obtained using the high-resolution echelle E140M grating of the Space Telescope
Imaging Spectrograph (STIS) in 2000-2001 to monitor the UV variability of NGC\,3783 and to
study its intrinsic absorption (Gabel et al.\,2003a, 2003b), but these spectacular STIS spectra never
were used to confirm and extend the earlier abundance results in the LA or to use the full
diagnostic power of the combined STIS/FUSE/ATCA data set to constrain the physical
conditions in the gas. An update of the earlier LA abudance measurements for the NGC\,3783
sightline is also highly desired to better characterize the dust-depletion properties
of the gas in the LA and to account for the change in the solar reference abundances
for the elements sulfur and oxygen since 2002 (Asplund et al.\,2009).

With this paper, we are closing this gap and provide the most precise determination 
of gas and dust abundances in two individual clouds with the LA based 
on the archival E140M STIS data of NGC\,3783.
Combining the STIS, FUSE, and ATCA data we further determine important physical
quantities such as density, temperature, thermal pressure, and absorber size in the
LA. By combining our precision results for the NGC\,3783 sightline with our recent 
measurements of the large-scale metal distribution in the LA based on lower-resolution
UV data from the Cosmic Origins Spectrograph (COS; Fox et al.\,2018), 
our multi-instrument spectral survey provides important new constraints on the 
origin of the LA, its physical properties, and its location in the Milky Way halo. 

%

\section{Observations and data analysis}

\subsection{HST/STIS data}

The archival high-resolution STIS data of NGC\,3783 ($l=287.5, b=+22.9$)
were taken in 2000 and 2001 using the E140M grating  
as part of the HST proposals 8029 (PI: Kraemer)
and 8606 (PI: Crenshaw). The STIS data have a spectral resolution
of $R\approx 45,000$, which corresponds to a velocity resolution
of $\Delta v \approx 6.6$ km\,s$^{-1}$ (FWHM). The signal-to-noise (S/N) in the
data is generally very high, reaching up to $\sim 45$ per resolution element
near 1255 \AA. 
This makes this archival spectrum an excellent data set to resolve 
the LA's internal velocity-component structure and determine the 
chemical composition of the gas towards NGC\,3783 at high accuracy.
 
The STIS data were originally 
reduced with the CALSTIS reduction pipeline (Brown et al.\,2002) as part of our earlier 
study of weak O\,{\sc i} absorbers/Lyman-limit systems in the Milky Way 
halo (Richter et al.\,2009), following the strategy outlined in Naranayan et al.\,(2005).
The STIS spectral data of the LA absorption towards NGC\,3783 has also been
shortly presented in our STIS HVC absorption survey (Herenz et al.\,2013), but no
detailed analysis of the LA metal abundances based on this superb 
UV data set has been published so far. 

\subsection{FUSE data}

The FUSE data were obtained in 2001 and 2004 (FUSE program IDs E031 and B107;
PI: Crenshaw for both proposals) with a total exposure time of 196 ks
in the various LiF and SiC spectral channels.
The FUSE spectra have a resolution of $\sim 20$ km\,s$^{-1}$
(FWHM), and cover the wavelength range $912-1180$ \AA.
The FUSE data were reduced with the CALFUSE reduction pipeline
(v.3.2.1). Individual spectra were coadded after correcting for
wavelengths shifts that are identifid by using interstellar lines as wavelength
reference. The final S/N per resolution element ranges between
$5$ and $30$, depending on wavelength and spectral channel. It is $24$
near 1100 \AA\, (LiF2 channel), where the most important H$_2$ lines 
for the analysis of the LA absorption are located.

Note that an analysis of H$_2$ absorption in the LA towards NGC\,3783 has published earlier by
Sembach et al.\,(2001) and Wakker (2006). While the Sembach et al.\,(2001)
study was based on a small subset of the FUSE data of NGC\,3783 available today, the Wakker (2006) 
study covers the full FUSE data set as analyzed here. Our motivation for re-analyzing the 
H$_2$ absorption in the FUSE spectral data is that with the high-resolution STIS data and the 
detection of neutral carbon (C\,{\sc i}) in the LA there is new information on the
Doppler parameter ($b$ value) in the H$_2$-bearing gas that affects the determination
of the H$_2$ column densities and rotational excitation temperature (see Sect.\,4.1).

We complement our UV data set for NGC\,3783 from STIS and FUSE with medium resolution 
(FWHM$\sim 20$ km\,s$^{-1}$) optical data from the Anglo-Australian Telescope (AAT;
as presented in West et al.\,1985), which covers the spectral region near
the Ca\,{\sc ii} H\&K lines at $\lambda 3930-3980$ \AA.


\begin{figure}[t!]
\epsscale{1.10}
\plotone{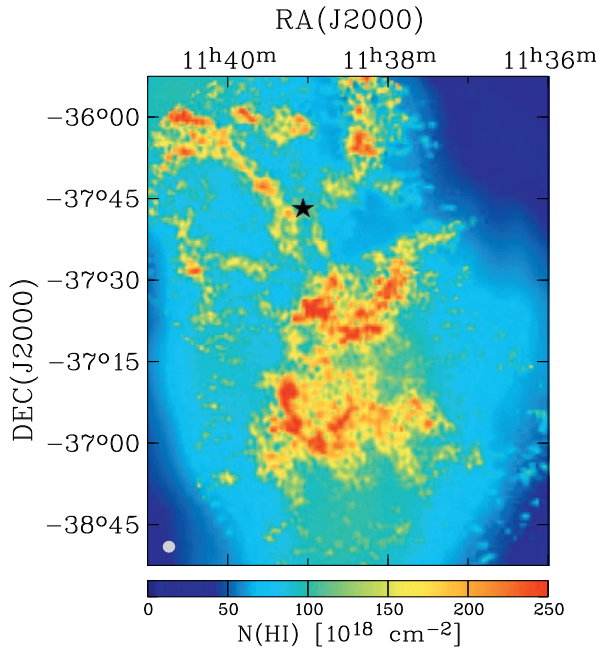}
\caption{H\,{\sc i} column-density map of high-velocity gas 
at $v_{\rm LSR}=220-260$ km\,s$^{-1}$ in the LA\,II region,
based on 21\,cm emission data from the GB140\,ft telescope and the ATCA 
(see Sect.\,2.3). The black star symbol marks the position of NGC\,3783. 
The effective beam
size is indicated with the gray circle in the lower left corner.}
\end{figure}


\begin{figure*}[t!]
\epsscale{1.00}
\plotone{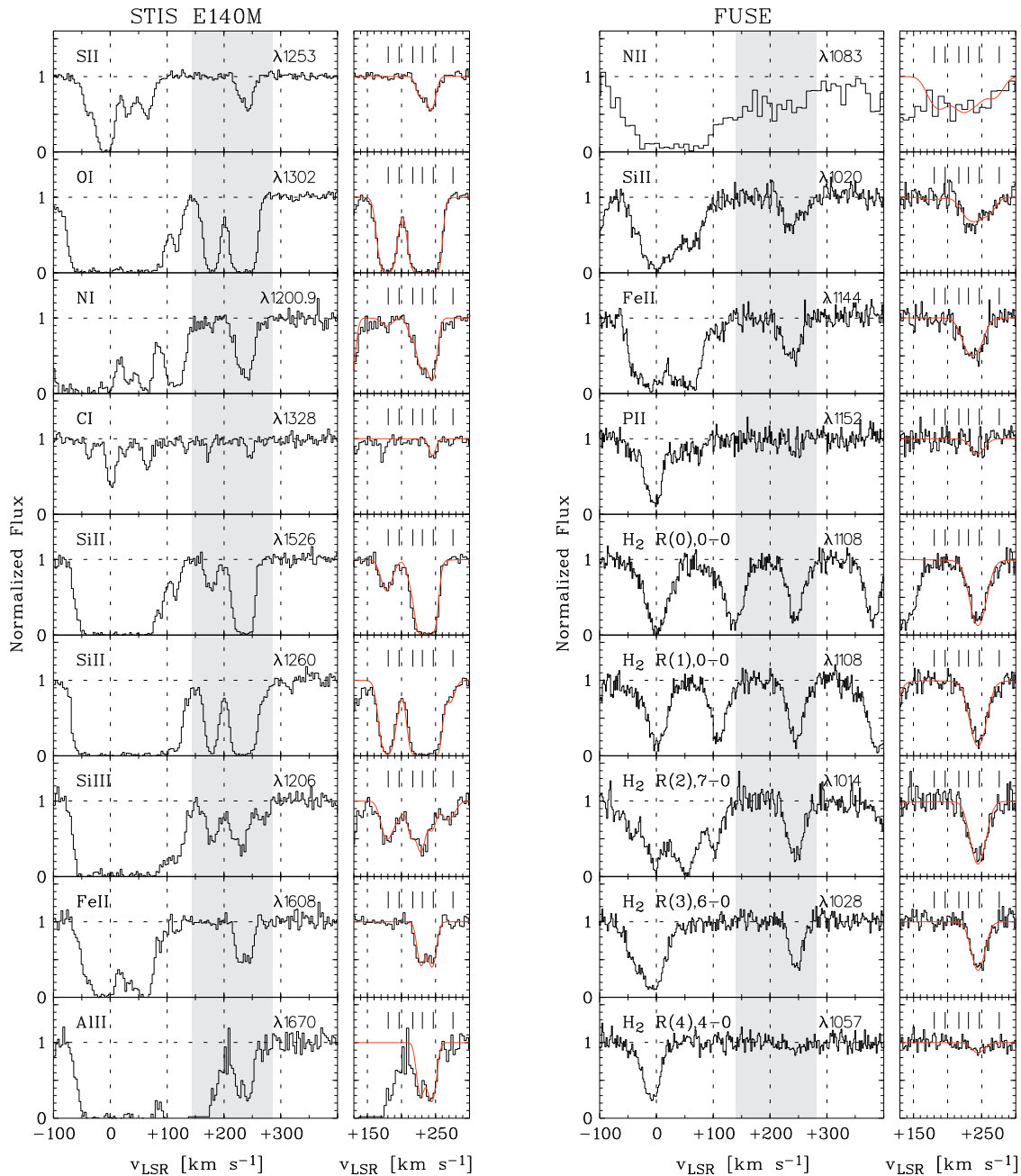}
\caption{Velocity profiles of various metal ions and H$_2$
({\it Left column:} STIS E140M data; {\it Right column:}
FUSE data). The gray-shaded area indicates the LA velocity
range. The red solid line shows the best-fitting result from
the component modeling with the black tick-marks displaying the
six identified LA velocity components. We do not show other 
available metal transitions that are either blended with
other absorption features or too weak too show significant 
absorption in LA\,II.}
\end{figure*}


\begin{figure}[t!]
\epsscale{1.00}
\plotone{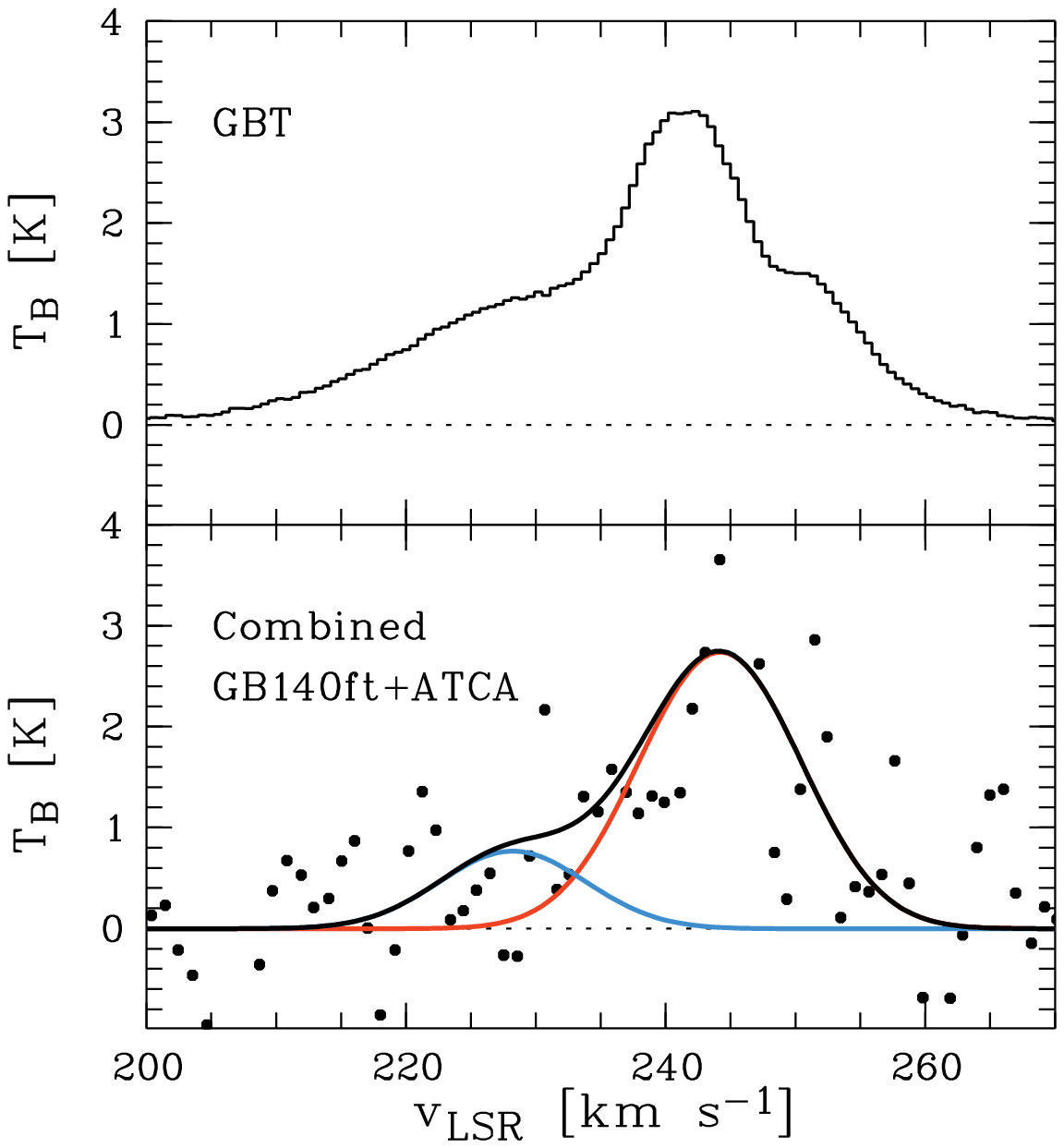}
\caption{
{\it Upper panel:}
H\,{\sc i} 21\,cm emission spectrum in the NGC\,3783 direction from the GBT (at 
$9.1 \arcmin$ resolution) in the velocity range $v_{\rm LSR}=200-270$ km\,s$^{-1}$.
{\it Lower panel:} combined ATCA/GB140\,ft data set (at $1 \arcmin$ resolution; 
black dots) from WOP02 for the same sightline/velocity range,
overlaid with a two-component fit (black solid line) and two fitted components
as individual profiles (red and blue solid lines).
}
\end{figure}


\subsection{H\,{\sc i} 21\,cm data}

For our study of the NGC\,3783 sightline, we make use of the 21\,cm interferometer 
data from ATCA presented in WOP02, which have a high angular
resolution of $1 \arcmin$. These high-resolution data were combined 
with single-dish 21\,cm data from the Green Bank 140\,ft telescope (GB140\,ft; 
$21 \arcmin$ angular resolution) and the Parkes Telescope ($17 \arcmin$
angular resolution) to construct the final 21\,cm spectrum in the 
NGC\,3783 direction (see WOP02; their Sect.\,4).

As part of our initial LA metallicity survey (Fox et al.\,2018) we
also obtained new high-sensitivity 21\,cm data for the 
NGC\,3783 sightline from the Robert C. Byrd Green
Bank Telescope (GBT; program IDs: GBT12A 206 \& GBT17B 424),
which provides an angular resolution of $9.1 \arcmin$ at
1420 MHz.
The GBT data cover the LSR velocity range between $-450$ and $+550$
km\,s$^{-1}$ at velocity resolution of $0.15$ km\,s$^{-1}$
and were obtained using frequency switching. For the basic
data reduction, we followed the procedures outlined in 
Boothroyd et al.\,(2011). The rms noise in the GBT spectrum
of NGC\,3783 is only 11 mK per 0.6 km\,s$^{-1}$ wide channel,
which allows us measure H\,{\sc i} column at levels 
$\leq 3\times 10^{18}$ cm$^{-2}$.


\subsection{Spectral analysis}

For the determination of the metal abundances in LA\,II, we use the following  
(unblended) ion transitions that are available in the combined STIS and FUSE data set:
C\,{\sc i} $\lambda\lambda 1277.2,1280.1,1328.8,1560.3,1656.9$,
N\,{\sc i} $\lambda 1200.7$,
N\,{\sc ii} $\lambda 1084.0$,
O\,{\sc i} $\lambda 1302.2$,
Al\,{\sc ii} $\lambda 1670.8$,
Si\,{\sc ii} $\lambda\lambda 1190.4,1193.3,1260.4,1304.4,1526.7$,
Si\,{\sc iii} $\lambda 1206.5$,
P\,{\sc ii} $\lambda 1152.8$,
S\,{\sc ii}  $\lambda 1253.8$,
and Fe\,{\sc ii} $\lambda\lambda 1143.2,1144.9,1608.5$.
Other available transitions in the STIS/FUSE wavelength ranges from these and other 
ions are either too weak or blended with other absorption lines (e.g., from intervening
systems), so that they cannot be used to study absorption in the LA.
Atomic data (laboratory wavelengths and oscillator strengths) have been adopted
from Morton (2003). For the analysis of the H$_2$ spectrum, we make use of the H$_2$ 
line-compilation from Abgrall \& Roueff (1989).

Column densities for the above-listed metal ions have been derived by Voigt-profile fitting
and component modeling, as described in detail in R13.
In addition, we use the apparent-optical-depth (AOD) method (Savage \& Sembach 1991) to measure
integrated column densities and to provide a consistency check with the
fitting/modeling method. In addition to the column density, the Voigt-profile fitting
provides for each LA\,II absorption component the $b$ value.
The $b$ value in the inner, dense core of LA\,II, as traced by C\,{\sc i}, was used
as reference value for the modeling of the H$_2$ absorption in the 
lower-resolution FUSE data, as C\,{\sc i} and H$_2$ trace the same gas phase
(e.g., Spitzer 1978). To validate the results of the H$_2$ modeling we also construct a 
curve of growth for blend-free H$_2$ absorption lines, as described in 
the Appendix.

Note that in this study, we do not re-analyze the absorption of higher ions 
(e.g., O\,{\sc vi}, C\,{\sc iv}, Si\,{\sc iv}) in the combined STIS/FUSE
data set, but concentrate instead on the metal abundances
and physical conditions in the weakly-ionized, cooler gas in the 
core of LA\,II.


\begin{figure}[t!]
\epsscale{1.00}
\plotone{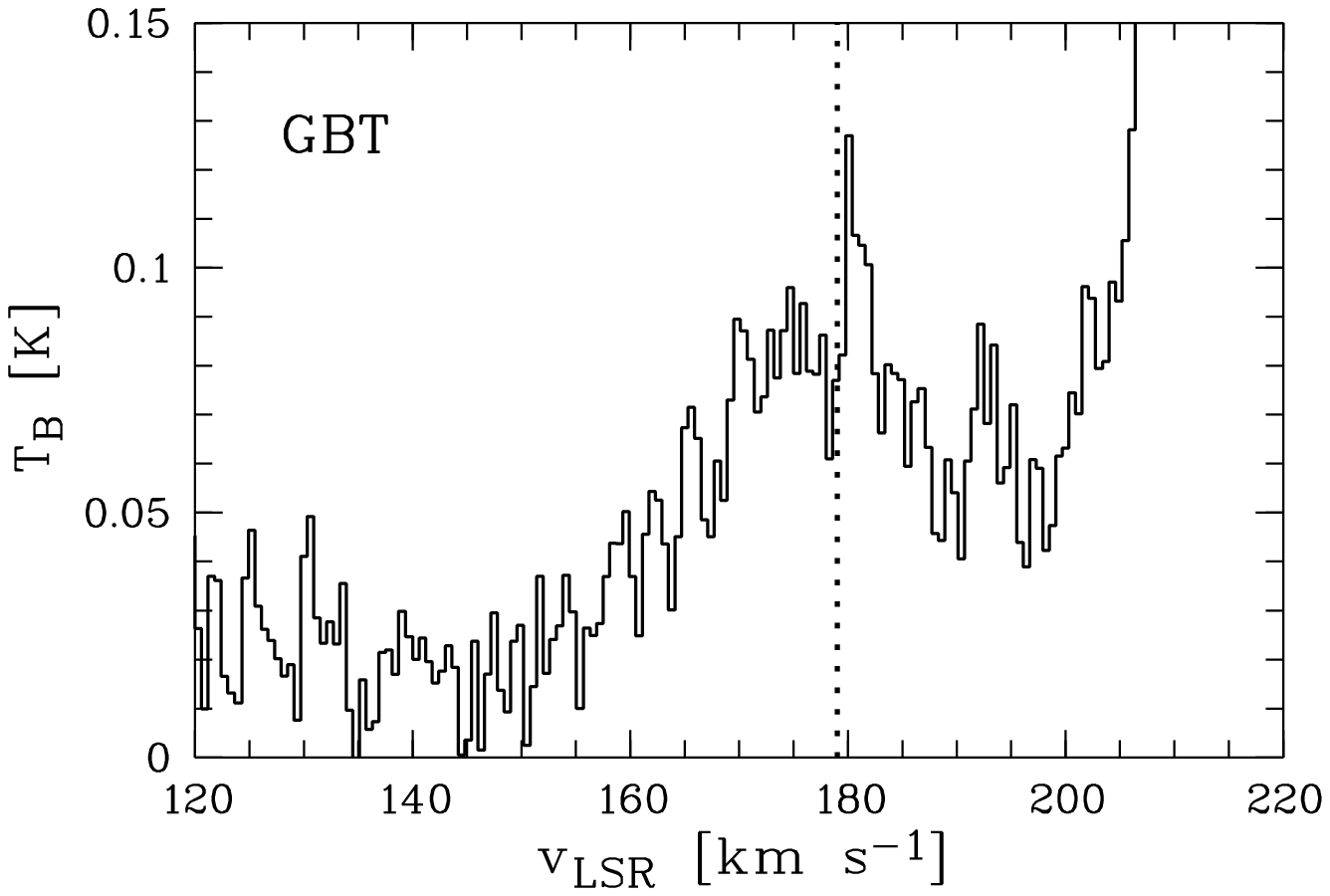}
\caption{
Zoom-in of the GBT 21\,cm spectrum towards NGC\,3783 in the velocity range 
$v_{\rm LSR}=120-220$ km\,s$^{-1}$. Very weak but significant emission
is seen in a broad component centered at $v_{\rm LSR}\approx 175-180$ km\,s$^{-1}$,
associated with the O\,{\sc i}/Si\,{\sc ii} absorption component at $+179$ km\,s$^{-1}$
(dashed line; see also Fig.\,2).
}
\end{figure}


\section{Structure and chemical composition of LA\,II}

\subsection{Component structure}

In Fig.\,1 we show the sky position of the NGC\,3783 sightline (black solid star
symbol) overlaid on the 21\,cm emission map of the LA gas distribution
from the combined ATCA/Parkes data (WOP02) in the velocity range 
$v_{\rm LSR}=220-260$ km\,s$^{-1}$. As discussed in WOP02, the
NGC\,3783 sightline does not pass through any of the peaks in the 21\,cm 
column-density distribution, but passes it at a column-density valley.
The observed angular fine-structure in the 21\,cm emission map at arcmin scale 
underlines that a precise estimate of the H\,{\sc i} column density in the LA 
along the NGC\,3783 sightline cannot be obtained from lower-resolution 
single-dish 21\,cm data alone.

In Fig.\,2 we show the velocity profiles of various metal ions and H$_2$
from the STIS E140M and FUSE data of NGC\,3783. The velocity range that is expected for
absorption from the LA ($v_{\rm LSR}=+140$ to $+300$ km\,s$^{-1}$; 
Putman et al.\,1998) is shown as gray-shaded area. In this range,
we identify six individual velocity components at 
$v_{\rm LSR}=+179,+195,+215,+229,+245,+269$ km\,s$^{-1}$.
The dominating neutral gas component (in terms of the neutral gas column density)
is located at $+245$ km\,s$^{-1}$ (comp.\,1),
coincinding in velocity with the dominating 21\,cm emission peak seen
in the 21\,cm data (Fig.\,3). This component is detected in absorption in O\,{\sc i},
N\,{\sc i} and in all singly-ionized species, as well in 
C\,{\sc i} and H$_2$. The simultaneous detection of neutral carbon and
molecular hydrogen at moderate strength already indicates that this component 
harbors predominantly neutral gas that is relatively cold and dense
(a phase that is referred to as ``cold neutral medium'', CNM).
The $+245$~km\,s$^{-1}$ component is accompanied by a weaker satellite
component at $+229$~km\,s$^{-1}$ (comp.\,2), which gives the unsaturated or only mildly saturated
LA absorption features of S\,{\sc ii}, N\,{\sc i}, Si\,{\sc iii}, P\,{\sc ii}, Fe\,{\sc ii}, 
and Al\,{\sc ii} at $+220-260$~km\,s$^{-1}$ an asymmetric shape (Fig.\,2).

Another major absorption component is seen at $+179$~km\,s$^{-1}$ (comp.\,3)
in the strong lines of O\,{\sc i} and Si\,{\sc ii} (also seen in Si\,{\sc iii}
and, very weak, in N\,{\sc i}), accompanied by its own satellite component
at $+195$~km\,s$^{-1}$ (comp.\,4). Two additional, very weak satellite components 
at $+215$ (comp.\,5) and $+269$~km\,s$^{-1}$ (comp.\,6) are visible only the strong 
lines of Si\,{\sc ii} and Si\,{\sc iii}. 

\subsection{H\,{\sc i} 21\,cm emission}

\subsubsection{Velocity range $200-280$ km\,s$^{-1}$}

In view of the complex velocity-component structure of LA\,II seen in the STIS data
we decided to re-fit also the combined ATCA 21\,cm data originally presented in WOP02.
In Fig.\,3 we compare the new $9.1 \arcmin$-resolution GBT spectrum for the 
NGC\,3783 direction (upper panel) with the $1 \arcmin$-resolution data from
ATCA (lower panel, black dots; from WOP02). The GBT data show a beam-smeared
version of the H\,{\sc i} distribution in the general direction of NGC\,3783,
with the main emission peak located at $v_{\rm LSR}=+241$ km\,s$^{-1}$,
an extended wing at lower velocities down to $\sim +210$ km\,s$^{-1}$,
and a weak bump at $\sim 250$ km\,s$^{-1}$. This velocity-component
structure slightly deviates from the one seen in absorption (see above).
This demonstrates that in this particular direction (Fig.\,1) 
small-scale structure exists in the 21\,cm distribution that is not fully
resolved in the $9.1 \arcmin$ beam of the GBT. This aspect
emphasizes the need for using interferometer data to determine an accurate
H\,{\sc i} column-density estimate in LA\,II towards NGC\,3783.

In WOP02, the combined $1 \arcmin$-resolution ATCA data for the NGC\,3783 direction  
was fitted with a single-component Gaussian
centered at $v_{\rm LSR}=+244$ km\,s$^{-1}$, yielding an H\,{\sc i} column
density of $N$(H\,{\sc i}$)=(8.26\pm 1.98) \times 10^{19}$ cm$^{-2}$
(see their Fig.\,9). However, the STIS data of NGC\,3783 demonstrates that
there are {\it two} main neutral components close together at $+245$ km\,s$^{-1}$ (comp.\,1) and 
$+229$ km\,s$^{-1}$ (comp.\,2; see Table 1). Also the GBT data 
indicate that neutral gas near $+229$ km\,s$^{-1}$ must be indeed present
on larger angular scales in the general NGC\,3783 direction. We therefore
re-fitted the combined ATCA data with a two-component fit, where
the center velocites were fixed to the absorption velocities of comp.\,1 and 2 at
$+245$ and $+229$ km\,s$^{-1}$. Also the FWHM of the two fitted emission components were fixed.
They were calculated from the $b$ values of comp.\,1 and 2. We thus are explicitly assuming that the 
$b$ values reflect only the internal velocity dispersion in the gas and 
thermal broadening is negligible. This is justified in view of the expected very
small contribution of thermal line-broadening to $b$ for metal lines
in cold neutral gas.

The fit to the raw ATCA data is shown in the lower panel of 
Fig.\,3 with the black solid line, while the red and blue solid lines
indicate the individual shapes of comp.\,1 and 2, respectively.
This two-component fit yields column densities of 
$N$(H\,{\sc i}$)=7.71\times 10^{19}$ cm$^{-2}$ for comp.\,1 and
$N$(H\,{\sc i}$)=1.94\times 10^{19}$ cm$^{-2}$ for comp.\,2.
The total H\,{\sc i} column density then comes out to
$N$(H\,{\sc i}$)=(9.65\pm 1.67) \times 10^{19}$ cm$^{-2}$
or log $N$(H\,{\sc i}$)=19.98^{+0.07}_{-0.08}$. This 
is $\sim 16$ percent higher than the value given in WOP02, but
both values formally agree at the $1\sigma$ level.
To verify the robustness of the fit and to account for the 
large scatter in the noisy $T_{\rm B}$ distribution
we also binned the ATCA data applying different bin 
sizes (ranging from 3 to 6 pixels) and different binning-starting 
points. However, we do not find any significant deviations from
the fit to the raw data.

We note that for $T\approx 100$ K and $b\approx 7$ km\,s$^{-1}$ 
the opacity in the 21\,cm line is $\leq 4$ percent in comp.\,1 and 2
(see Dickey \& Lockman 1990) and much smaller in all other 
components. Therefore, any possible corrections related to the 21\,cm line
opacity are negligible compared to the 17 percent uncertainty that we quote 
on $N$(H\,{\sc i}) from the fit of the ATCA data.

In the following abundance determination, we therefore use the value 
log $N$(H\,{\sc i}$)=19.98^{+0.07}_{-0.08}$ for the total neutral gas column density 
in comp.\,1+2.


\begin{deluxetable*}{llrrrrrrrrr}
\tabletypesize{\scriptsize}
\tablewidth{0pt}
\tablecaption{Summary of column-density measurements}
\tablehead{
\colhead{} &
\colhead{Component\tablenotemark{a}} & \colhead{1} & \colhead{2} & \colhead{3} & \colhead{4} & \colhead{5} & \colhead{6} & & \\
\colhead{} &
\colhead{$v$ [km\,s$^{-1}$]} & \colhead{$+245$} & \colhead{$+229$} & \colhead{$+179$} & \colhead{$+195$} &
\colhead{$+215$} & \colhead{$+269$} &  &  & \\
\colhead{} &
\colhead{$b$ [km\,s$^{-1}$]} & \colhead{$7.3$ ($5.0$\tablenotemark{b})} & \colhead{$6.6$} & \colhead{$10.0$} & \colhead{$9.0$} &
\colhead{$6.5$} & \colhead{$12.0$} & & & \\
\\
\colhead{Ion/Species} &
\colhead{Instrument} & \colhead{log\,$N_1$} & \colhead{log\,$N_2$} & \colhead{log\,$N_3$} & \colhead{log\,$N_4$} &
\colhead{log\,$N_5$} & \colhead{log\,$N_6$} & \colhead{log\,$N_{\rm 1+2,model}$} & 
\colhead{log\,$N_{\rm 1+2,AOD}$\tablenotemark{c}} & \colhead{log\,$N_{\rm 3,AOD}$\tablenotemark{c}} \\
} 
\startdata
H\,{\sc i}           & ATCA+GB140ft & ...  & ...   & ...   & ...   & ...   & ...   &  ...              & $19.98^{+0.07}_{-0.08}$ & ... \\
                     & GBT         & ...   & ...   & ...   & ...   & ...   & ...   & ...               & ...              & $18.69^{+0.11}_{-0.15}$ \\    
C\,{\sc i}           & STIS        & 13.00 & 12.80 & ...   & ...   & ...   & ...   & $13.05 \pm 0.04$  & $13.03 \pm 0.04$ & ... \\
C\,{\sc i}$^{\star}$ & STIS        & ...   & ...   & ...   & ...   & ...   & ...   &  ...              & $\leq 12.75$     & ... \\
C\,{\sc i}$^{\star\star}$ & STIS   & ...   & ...   & ...   & ...   & ...   & ...   &  ...              & $\leq 12.40$     & ... \\
C\,{\sc ii}$^{\star}$ & STIS       & ...   & ...   & ...   & ...   & ...   & ...   &  ...              & $\leq 12.93$     & ... \\
N\,{\sc i}           & STIS        & 14.30 & 14.05 & 13.20 &   ... & 13.43 & ...   & $14.49 \pm 0.06$  & $14.46 \pm 0.06$ & $13.14 \pm 0.10$ \\
N\,{\sc ii}          & FUSE        & 13.39 & 13.51 & 13.50 & 13.50 & 13.46 & 13.50 & $13.72 \pm 0.09$  & $13.84 \pm 0.11$ & ... \\
O\,{\sc i}           & STIS        & 15.90 & 15.60 & 14.90 & 13.30 & 14.20 & ...   & $16.08 \pm 0.64$  & $\geq 15.14$     & $\geq 14.69$ \\
Al\,{\sc ii}         & STIS        & 12.45 & 12.31 & ...   & ...   & ...   & ...   & $12.69 \pm 0.06$  & $12.72 \pm 0.06$ & ... \\
Si\,{\sc ii}         & STIS+FUSE   & 14.19 & 14.14 & 13.28 & 11.84 & 12.98 & 11.48 & $14.61 \pm 0.05$  & $14.55 \pm 0.08$ & $13.24 \pm 0.08$ \\
Si\,{\sc iii}        & STIS        & 12.14 & 12.48 & 12.41 & 11.90 & 12.23 & 12.11 & $12.62 \pm 0.07$  & $12.57 \pm 0.09$ & $12.49 \pm 0.06$ \\
P\,{\sc ii}          & FUSE        & 12.90 & ...   & ...   & ...   & ...   & ...   & $12.90 \pm 0.06$  & $12.73 \pm 0.13$ & ... \\
S\,{\sc ii}          & STIS        & 14.40 & 14.08 & ...   & ...   & ...   & ...   & $14.57 \pm 0.03$  & $14.59 \pm 0.03$ & $\leq 13.40$ \\
Fe\,{\sc ii}         & STIS+FUSE   & 13.72 & 13.66 & ...   & ...   & ...   & ...   & $13.99 \pm 0.06$  & $13.99 \pm 0.05$ & $\leq 12.80$ \\
Ca\,{\sc ii}         & AAT         &   ... & ...   & ...   & ...   & ...   & ...   &  ...              & $11.76^{+0.11}_{-0.15}$ & ... \\
\\
H$_2, J=0$           & FUSE        & 18.00 & ...   & ...   & ...   & ...   & ...   & $18.0 \pm 0.4$    & ... & ... \\
H$_2, J=1$           & FUSE        & 17.80 & ...   & ...   & ...   & ...   & ...   & $17.8 \pm 0.4$    & ... & ... \\
H$_2, J=2$           & FUSE        & 16.40 & ...   & ...   & ...   & ...   & ...   & $16.4 \pm 0.2$    & ... & ... \\
H$_2, J=3$           & FUSE        & 15.20 & ...   & ...   & ...   & ...   & ...   & $15.2 \pm 0.2$    & ... & ... \\
H$_2, J=4$           & FUSE        & 14.00 & ...   & ...   & ...   & ...   & ...   & $14.0 \pm 0.2$    & ... & ... \\
H$_2$, total         & FUSE        & 18.22 & ...   & ...   & ...   & ...   & ...   & $18.2 \pm 0.4$    & ... & ... \\
\enddata
\tablenotetext{a}{Velocity centroids and $b$-values are kepted fix in the component model; see Sect.\,3.3.}
\tablenotetext{b}{Only for C\,{\sc i} and H$_2$.}
\tablenotetext{c}{Transitions used to dermine column desities via the AOD method:
H\,{\sc i} 21cm, C\,{\sc i} $\lambda\lambda 1328,1377$, N\,{\sc i} $\lambda\lambda 1134,1200$, N\,{\sc ii} $\lambda 1083$, O\,{\sc i} $\lambda 1302$,
Al\,{\sc ii} $\lambda 1670$, Si\,{\sc ii} $\lambda\lambda 1193,1020$, Si\,{\sc iii} $\lambda 1206$, P\,{\sc ii} $\lambda 1152$, S\,{\sc ii} $\lambda 1253$,
Fe\,{\sc ii} $\lambda\lambda 1144,1608$. The adopted velocity range is $200-280$ km\,s$^{-1}$.}
\end{deluxetable*}


\subsubsection{Velocity range $150-200$ km\,s$^{-1}$}

Because of the very high quality of the GBT data (noise level is
$11$ mK per $0.6$ km\,s$^{-1}$ wide channel) the GBT spectrum
allows us to search for weak emission from neutral gas that is 
associated with the $+179$ km\,s$^{-1}$ absorption component (comp.\,3) seen in
O\,{\sc i}, N\,{\sc i}, Si\,{\sc ii}, and Si\,{\sc iii} (see Table 1).

In Fig.\,4 we show a zoom-in into the GBT spectrum in the velocity 
range between $+120$ and $+220$ km\,s$^{-1}$ for 
$T_{\rm B}\leq 150$ mK. A broad, weak emission peak 
is clearly visible near $+180$ km\,s$^{-1}$. Integration of 
$T_{\rm B}$ over the velocity range $+155$ to $+215$ km\,s$^{-1}$ yields
a column density of $N$(H\,{\sc i}$)=(4.70\pm 1.35) \times 10^{18}$ cm$^{-2}$
or log $N$(H\,{\sc i}$)=18.67^{+0.11}_{-0.15}$. The neutral 
gas-column density of the component at $+179$ km\,s$^{-1}$ thus
is $\sim 20$ times smaller than $N$(H\,{\sc i}) in the main
LA\,II components near $+240$ km\,s$^{-1}$.

\subsection{Absorption-profile fitting and modeling}

For the quantitative analysis of the observed absorption pattern, we reconstructed
the component structure by combining Voigt-profile fitting and component
modeling (see also R13). For this, we simultaneously fitted the 
(relatively weak absorption) lines of 
S\,{\sc ii} $\lambda 1253.8$, 
N\,{\sc i} $\lambda 1200.7$,
Si\,{\sc iii} $\lambda 1206.5$,
Al\,{\sc ii} $\lambda 1670.8$,
and Fe\,{\sc ii} $\lambda 1608.5$
in the STIS data to determine the
center velocities, $b$-values, and column densities, $N$,
for these ions in comp.\,1 and 2. We then fixed the
velocities and $b$-values for these two components and
fitted (as the only free parameter) the column densities 
for the remaining two ions that show only strongly saturated
absorption in these components (O\,{\sc i} and Si\,{\sc ii}).
For the remaining four absorption components, we then obtained 
center velocities, $b$-values, and column densities 
by simultaneously fitting those (strong) lines in which 
these (weak) components show up. In this way, we have constructed 
a final component model for which we have generated a synthetic
spectrum taking into account the spectral resolution of the STIS
and FUSE instruments. This synthetic spectrum is shown in Fig.\,2
with the red solid line. $1\sigma$ column-density errors have been
derived by varying $N$ in the model in the allowed range, as constrained
by the residuals between the model spectrum and the data (see also
R13, their App.\,B).
All results from the fitting/modeling procedure are summarized in Table 1.

As it turns out, all metal ions except C\,{\sc i} can be fitted
by assigning a fixed center velocity and a fixed $b$-value to each
component. The five C\,{\sc i} lines at 
$\lambda\lambda 1277.2,1280.1,1328.8,1560.3,1656.9$ \AA,
that show absorption in the LA in comp.\,1 at 
$+245$ km\,s$^{-1}$ are too narrow to be reproduced by the
$b$-value of $7.3$ km\,s$^{-1}$ that is derived from the fitting/modeling
of the singly-ionized species and O\,{\sc i}/N\,{\sc i}. If we fit 
the C\,{\sc i} lines individually, we obtain a $b$-value of 
$(5.0\pm 0.6)$ km\,s$^{-1}$. The smaller $b$-value for C\,{\sc i} 
indicates that the neutral carbon resides in sub-structure
within comp.\,1 that is spatially more confined than the
region that hosts O\,{\sc i}/N\,{\sc i} and the singly ionized species,
reflected by a smaller velocity dispersion in the gas. 
This trend is not surprising, however. The ionization potential of 
neutral carbon is only $11.26$ eV and therefore substantial 
amounts of C\,{\sc i} can exist only in relatively dense regions
where the recombination rate is sufficiently high to compensate
for the photoionization. We conclude that the C\,{\sc i} absorption 
arises in a small, dense gas clump within comp.\,1 that is surrounded 
by more diffuse gas (the latter phase in comp.\,1 also being traced by the
other ions that have higher ionization potentials).

The fact that we derive a $b$-value for C\,{\sc i} in comp.\,1 that is 
lower than for the other ions has implications for the absorption-line
modeling of the LA molecular hydrogen absorption in the FUSE data.
From previous UV absorption-line studies it is well known that neutral
carbon co-exists with H$_2$ in the same gas phase (e.g., Spitzer 1978),
so that $b$(H$_2)=b$(C\,{\sc i}$)=(5.0\pm 0.6)$ km\,s$^{-1}$ is expected.
This value is very similar to the value derived in Wakker (2006).
In the Appendix, we provide another, independent measurement of
$b$(H$_2$) from the curve-of-growth analysis of selected H$_2$
lines from different rotational levels.

For Ca\,{\sc ii} $\lambda 3934$, we adopt the equivalent-width
measurement in LA\,II from the 
AAT data, as presented in West et al.\,(1985). These authors
measure $W_{3934}=(50 \pm 15)$ m\AA\,in comp.\,1+2, which translates 
to a column density of log $N$(Ca\,{\sc ii}$)=11.76^{+0.11}_{-0.15}$.


\subsection{Metal abundances}

Based on the modeled ion column densities listed in Table 1 we obtain
the relative gas-phase abundance for an element M compared to the
solar abundance in the standard way


\begin{equation}
{\rm [M/H]\,=\,log\,(M_{\rm X}/H\,I)}\,+{\rm IC(M_{\rm X})}-\,{\rm log\,(M/H)}_{\sun},
\end{equation}


where X is an ion of the element M, IC(M$_{\rm X}$) is the ionization
correction for that ion, and M/H$_{\sun}$ is the solar reference
abundance for M. The results and assumed parameters are summarized in Table 2.

As solar reference abundances we use the values listed in Asplund et al.\,(2009), 
which deviate (substantially for some elements) from those used 
in the earlier LA abundance 
measurements for the NGC\,3783 sightline. For example: Lu et al.\,(1998) used
a value of (S/H)$_{\sun}=-4.73$ dex from Anders \& Grevesse (1989) for their metallicity 
determination of LA\,II, while Asplund et al.\,(2009) gives (S/H)$_{\sun}=-4.88$ dex, 
thus 0.15 dex lower (see also Howk et al.\,2005).

\subsubsection{Ionization corrections}

Although most of the metal ions listed in Table 1 represent the dominant ionization states
in neutral interstellar gas (C\,{\sc i} and Ca\,{\sc ii} being the exception), their ionization potentials 
are different from that of H\,{\sc i} so that ionization corrections (ICs) generally need to be 
considered. The neutral-gas column density in comp.\,1+2 is almost $10^{20}$ cm$^{-2}$ 
so that here the ICs are expected to be small. For comp.\,3, which has log $N$(H\,{\sc i}$)=18.67$,
the ICs are expected to be more severe and strongly dependent on the local gas density.

To quantitatively estimate the ICs, we follow the strategy outlined in our earlier 
MS/LA studies (Fox et al.\,2013, 2014, 2018; R13). From the photoionization 
modeling with the {\it Cloudy} code (Ferland et al.\,2013), we calculate the IC for each ion.
This calculation is
based on the local neutral gas column density in comp.\,1+2 and comp.\,3 and an estimate
of the local ionization parameter, $U$, which is the ratio between ionizing photon density and 
the gas density in the cloud. The local photon density (in units cm$^{-3}$) in LA\,II is set to log $n_{\gamma}=-5.55$.
This value is adopted from the 3-D Galactic radiation field model of Fox et al.\,(2014),
which takes into account the direction of the NGC\,3783 sightline 
and an assumed distance of LA\,II of $20$ kpc. 

The local value for $U$ can estimated from the measured Si\,{\sc iii}/Si\,{\sc ii} column density 
ratios, an approach that has been used in many of our previous studies of the MS and the LA
(see Fox et al.\,2013, 2014, 2018; R13).  One drawback of this method is
that it does not account for the possible multi-phase nature of the absorbers that may be composed
of denser cores and more diffuse gas layers where the local Si\,{\sc iii}/Si\,{\sc ii} ratios 
may vary considerably. In predominantly ionized gas layers, for example, 
Si\,{\sc iii} will dominate over Si\,{\sc ii}, but most of the total Si\,{\sc ii} column density
comes from the densest regions. This is aspect is relevant particularly for low-column
density CGM absorbers, where Si\,{\sc ii}/Si\,{\sc iii}/Si\,{\sc iv} often are believed 
to trace different gas phases. Evidence for this comes from the fact that for many absorbers 
a single-component {\it Cloudy} model cannot reproduce the observed 
column-density ratios of these ions (Muzahid et al.\,2018; Richter et al.\,2016), indicating
that they are likely not be well mixed.
Therefore, the absolute values for the ICs derived from the observed Si\,{\sc iii}/Si\,{\sc ii} 
column-density ratios, which represent sightline-averages per definition, should be regarded as
upper limits. They add to the total error budget for determining [M/H] as a 
result of the unknown density structure within the absorbers.

\subsubsection{Components 1 \& 2}

From the observed Si\,{\sc iii}/Si\,{\sc ii} column density ratio in comp.\,1+2 (Table 1, column 9),
we infer a value of the ionization parameter of log $U=-3.4$. The resulting ICs for each ion
in comp.\,1+2 are listed in the sixth column of Table 2. At this high neutral gas column density, 
the ICs are all very small, as expected (their absolute values are $\leq 0.03$ dex for N\,{\sc i}, 
O\,{\sc i}, Al\,{\sc ii}, P\,{\sc ii}, and S\,{\sc ii}) and $0.09$ and $0.12$ dex for Fe\,{\sc ii} 
and Si\,{\sc ii}, respectively). These numbers demonstrate that ionization effects have only a very 
minor influence on the determination of metal abundances in comp.\,1+2.
The resulting values for [M/H], derived by using equation (1), are listed in the seventh column of Table 2.

The well-resolved, unsaturated S\,{\sc ii} $\lambda 1253.8$ absorption in this STIS spectrum
has a very high S/N of 35 per resolution element. This line provides the most precise 
constraint of the $\alpha$ abundance in LA\,II. Unfortunately, the
two other available S\,{\sc ii} lines at $1250.6$ and $1259.5$ \AA\, are blended
with other absorption features. At the given H\,{\sc i} 
column density in comp.\,1+2, the IC for S\,{\sc ii} is zero (see Table 2), minimizing
the systematic error for determining [S/H]. In addition, sulfur is not depleted into
dust grains (such as the $\alpha$ element Si; Savage \& Sembach 1996). Our measurement
of log $N$(S\,{\sc ii})$=14.57\pm 0.03$ in comp.\,1+2 translates into a sulfur 
abundance of [S/H$]=-0.53\pm 0.08$ ($0.30\pm 0.05$ solar). This value is
$0.07$ dex higher than the previous estimate from Lu et al.\,(1998), but 
most of this discrepancy is due to the use of different solar reference
abundances (see above) and different values for $N$(H\,{\sc i}), while
the values for $N$(S\,{\sc ii}) are identical.
The value of [S/H$]=-0.53\pm 0.08$ in comp.\,1+2 from the STIS data is higher than the 
value derived by us from the lower-resolution COS data (Fox et al.\,2018;
[S/H$]=-0.63\pm 0.16$), but is consistent within the 1$\sigma$ error range. A more detailed comparison 
between the STIS and the COS data is presented in the Appendix.

Also for O\,{\sc i} and P\,{\sc ii} ionization corrections and dust depletion
effects are unimportant for the determination of [$\alpha$/H], but the 
O\,{\sc i} $\lambda 1302.2$ line is strongly saturated (Fig.\,1), so that 
log $N$(O\,{\sc i}) is uncertain (Table 1). The P\,{\sc ii} $\lambda 1152.8$
line lies in the FUSE wavelength range and is unsaturated, but has a S/N of only 12 
per resolution element, thus substantially lower than the 
S\,{\sc ii} $\lambda 1253.8$ line. Still, the derived abundances of 
[O/H$]=-0.60\pm 0.27$ and [P/H$]=-0.48\pm 0.09$ agree very well within
their error range with the above given S abundance. Therefore, the 
observed abundance of phosphorus provides additional strong contraints 
on the $\alpha$ abundance in LA\,II towards NGC\,3783.
We discuss elements that are known to be depleted into dust grains 
(e.g., Si, Fe, Al, and Ca; Savage \& Sembach 1996) in Sect.\,3.6.

For nitrogen, we derive [N/H$]=-1.31\pm 0.09$ in comp.\,1+2, thus substantially lower than
for $\alpha$ elements. Note that the lower abundance of N cannot be explained by an 
ionization effect. The {\it Cloudy} model indicates that the IC for N\,{\sc i} in comp.\,1+2
is only $-0.03$ dex, in line with the observed low column density of N\,{\sc ii}, which
is 0.75 dex below that of N\,{\sc i}. This is the first measurement of the nitrogen abundance in 
the LA. A similarly low value has been found in the main body of 
MS (Fox et al.\,2013) and in the MS-LMC filament (R13).
The resulting [N/$\alpha$] ratio in LA\,II is $-0.77$ dex, a value that is 
also typical for other low-metallicity MW halo clouds (e.g., Richter et al.\,2001)
and DLAs at low and high redshift (e.g., Pettini 2008; Petitjean et al.\,2008).
A low [N/$\alpha$] ratio is typical for environments that are
enriched predominantly by SNe Type II (see discussion in R13).


\begin{figure}[t!]
\epsscale{1.20}
\plotone{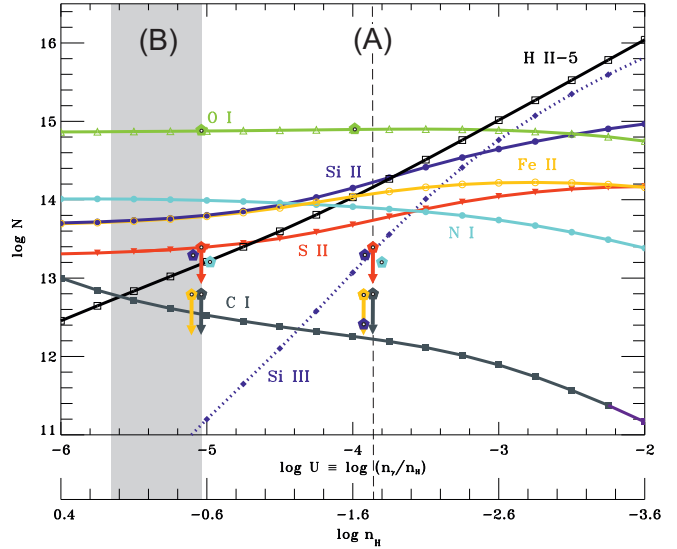}
\caption{
Predicted ion column densities vs. ionization parameter ($U$) and
gas density ($n_{\rm H}$) from the {\it Cloudy} model of comp.\,3. The
model-input parameters are: log $N$(H\,{\sc i}$=18.67$, [M/H$]=-0.53$,
log $n_{\gamma}=-5.5$. The two possible model solutions, (A) and (B), 
are discussed in Sect.\,3.4.3.
}
\end{figure}


\subsubsection{Component 3}

From our Cloudy modeling of comp.\,3, we find that the ionization correction 
for O\,{\sc i} is negligible for gas densities of log $n_{\rm H}\geq -3$, which 
is not surprising as both hydrogen and oxygen have the nearly identical first 
ionization potentials.
Therefore, the O\,{\sc i}/H\,{\sc i} ratio provides a reliable measure for the 
oxygen abundance also in comp.\,3. From our measurements listed in
Table 1 we derive [O/H$]=-0.46\pm 0.24$, which is (within the error
range) identical to the value derived for comp.\,1+2. The large 
error in this abundance estimate is due to the saturation of the
O\,{\sc i} $\lambda 1302$ line (Fig.\,2).

For the determination of gas-phase abundances (or upper limits) 
for the other elements (N, Si, C, Fe, S) we have set
up another {\it Cloudy} model assuming log $N$(H\,{\sc i}$)=18.67$
and an overall metallicity of $0.3$ solar. The {\it Cloudy} model plot, 
which shows the predictions for ion column densities as a function
of ionization parameter $U$, is shown in Fig.\,5.
If we take the observed Si\,{\sc iii}/Si\,{\sc ii} in comp.\,3 as the basis
(Table 1), the {\it Cloudy} model delivers log $U=-3.86$ (log $n_{\rm H}=-1.74$) as
unique solution (hereafter referred to as Model (A); dashed line in Fig.\,5). 
At this value of $U$, however, the observed column densities
(or upper limits) for S\,{\sc ii} and Si\,{\sc ii} do not match the
expectations from the {\it Cloudy} model for a uniform $\alpha$ abundance
of $0.3$ solar. Instead, the sulfur abundance would neeed to be 
lowered to a value of $0.14$ solar and the Si abundance to a value of 
$0.03$ to make the {\it Cloudy} model with log $U=-3.86$ fit the data.
Depletion into dust grains could explain some of the predicted underabundance 
of Si (see next section), but not all of it, while S is not expected to
be depleted into dust at a significant level. Abundance anomalies for S
and other $\alpha$ elements have been found in intervening metal absorbers (e.g.,
Fox, Anne et al.\,2014; Richter et al.\,2005a), but not in Galactic halo clouds.
In view of the solar S/O ratio in comp.\,1+2 it would be difficult to
explain why S and Si are substantially underabundant compared to O just
in this particular region of the LA.

If, however, only a fraction of the observed Si\,{\sc iii} column density
in comp.\,3 is co-spatial with Si\,{\sc ii} (i.e., if the absorber is
multi-phase), then the value of log $U=-3.86$ represents only an upper
limit (see discussion in Sect.\,3.4.1) and other solutions for log $U$ 
and log $n_{\rm H}$ are allowed. If we ignore Si\,{\sc iii} and instead fix 
the S/O ratio in comp.\,3 to the solar value and further take into
account that C\,{\sc i} is not detected in comp.\,3, the {\it Cloudy} model
gives the ranges log $U=-5.65$ to $-5.05$ and log $n_{\rm H}=-0.55$ to $+0.05$ 
as allowed solutions (Model (B); gray-shaded area in Fig.\,5). 
In this range for $U$, the metal abundances
come out as [O/H$]=-0.53\pm 0.24$, [S/H$]\leq -0.52$, [N/H$]=-1.30\pm 0.14$, 
[Si/H$]=-1.02\pm 0.15$, and [Fe/H$]\leq -1.50$, thus (within the uncertainties)
identical to the abundances derived for comp.\,1+2.
In view of the fact that Si\,{\sc iii} is known to be non-cospatial in many
CGM absorbers (Richter et al.\,2016) and Model (B) does not require any significant
abundance anomalies within the family of $\alpha$ elements (O, S, Si), we
regard Model (B) as the far more realistic solution to the {\it Cloudy} output. 
In the following, we therefore concentrate on Model (B) and do not
any further consider Model (A). Model (B) 
implies that the $+179$ km\,s$^{-1}$ component represents a relatively
dense ($n_{\rm H}\geq 0.3$ cm$^{-3}$) clump that is well separated from 
the main body of LA\,II in velocity
(possibly also in space), but that has the same origin.


\begin{figure}[b!]
\epsscale{1.20}
\plotone{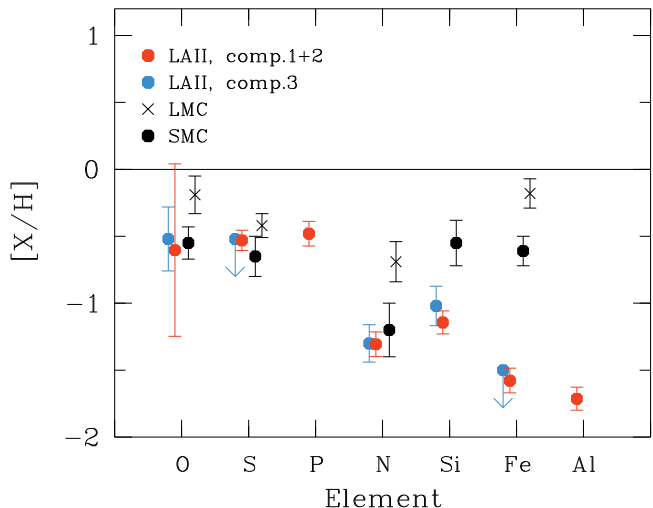}
\caption{
Gas-phase metal-abundances in LA\,II towards NGC\,3783, as derived from
our STIS/FUSE data set. The first three elements (O, S, P) represent $\alpha$ elements
that are not expected to be depleted into dust grains. The
following four elements are known to be underabundant in interstellar 
environments either because of nucleosynthetic effects (N, Fe) and/or 
due to dust depletion (Si, Fe, Al). 
Photospheric abundances in the LMC and SMC are shown for comparison
(adopted from Tchernyshyov et al.\,2015 and Jenkins \& Wallerstein 2017 and
references therein).
}
\end{figure}


\begin{deluxetable*}{lrrrrrrr}

\tablewidth{0pt}
\tablecaption{Summary of gas-phase abundances and dust depletion values in components 1+2}
\tablehead{
\colhead{Ion X of element M} & \colhead{$Z$} & \colhead{I.P.$^{\rm a}$} & 
\colhead{log\,(M/H)$_{\sun}^{\rm b}$} & 
\colhead{log\,(X/H\,{\sc i})$^{\rm c}$} & \colhead{I.C.$^{\rm d}$} & \colhead{[M/H]$^{\rm e}$} &
\colhead{log $\delta$(M)$^{\rm f}$} \\
\colhead{} & \colhead{} & \colhead{[eV]} & \colhead{$+12$} &
\colhead{} & \colhead{} & \colhead{} & \colhead{} \\
}
\startdata
C\,{\sc i}   &  6 & 11.26 & $8.43\pm0.05$   & $-6.93\pm0.07$  & ...     & ...            & ...     \\
N\,{\sc i}   &  7 & 14.53 & $7.83\pm0.05$   & $-5.49\pm0.09$  & $-0.03$ & $-1.31\pm0.09$ & ...     \\
N\,{\sc ii}  &  7 & 14.53 & $7.83\pm0.05$   & $-6.26\pm0.13$  & ...     & ...            & ...     \\
O\,{\sc i}   &  8 & 13.62 & $8.69\pm0.05$   & $-3.90\pm0.27$  & $-0.01$ & $-0.60\pm0.27$ & ...     \\
Al\,{\sc ii} & 13 & 18.83 & $6.45\pm0.03$   & $-7.29\pm0.09$  & $+0.03$ & $-1.71\pm0.09$ & $-1.18$ \\
Si\,{\sc ii} & 14 & 16.35 & $7.51\pm0.03$   & $-5.51\pm0.09$  & $-0.12$ & $-1.00\pm0.09$ & $-0.61$ \\
P\,{\sc ii}  & 15 & 19.77 & $5.41\pm0.04$   & $-7.08\pm0.09$  & $+0.01$ & $-0.48\pm0.09$ & ...     \\
S\,{\sc ii}  & 16 & 23.34 & $7.12\pm0.03$   & $-5.41\pm0.08$  &  $0.00$ & $-0.53\pm0.08$ & ...     \\
Ca\,{\sc ii} & 20 & 11.87 & $6.34\pm0.04$   & $-8.22\pm0.09$  & $\leq +1.35$ & $\geq -1.83$ & $\geq -1.35$ \\
Fe\,{\sc ii} & 26 & 16.12 & $7.50\pm0.04$   & $-5.99\pm0.09$  & $-0.09$ & $-1.58\pm0.09$ & $\geq -1.05$ \\

\enddata
\tablenotetext{a}{I.P.=\,ionization potential.} 
\tablenotetext{b}{Solar reference abundance for element M from Asplund et al.\,(2009).}
\tablenotetext{c}{Listed errors include uncertainties from the column-density
measurements of X and H\,{\sc i}.}
\tablenotetext{d}{I.C.=\,ionization correction, as defined in Sect.\,4.1.}
\tablenotetext{e}{Gas phase abundance for element M,
defined as [M/H]\,=\,log\,(M/H)\,$-$\,log\,(M/H)$_{\sun}$.}
\tablenotetext{f}{Depletion value, defined as 
log $\delta$(M)=\,[M/H$]_{\rm LA}-$[S/H$]_{\rm LA}$.} 

\end{deluxetable*}


\subsubsection{Comparison with LMC/SMC abundances}

In Fig.\,6 we compare the LA\,II abundances in comp.\,1+2 and comp.\,3 
from this study with photospheric abundances in the SMC and LMC
taken from the abundance compilations presented in Tchernyshyov et al.\,(2015) 
and Jenkins \& Wallerstein (2017; see references therein).
The similarity between the LA\,II abundances in comp.\,$1-3$ and the
SMC abundance pattern for non-depleted species (O, S, P, N) is striking.
Within the error bars, the abundances for these elements are identical
in LA\,II and in the stellar body of the SMC, while the LMC abundances
are all systematically higher. 

For the interpretation of Fig.\,6, it is important to note that the 
chemical composition of the LA\,II today should reflect the ISM abundance 
in the host galaxy at the time when LA\,II was separated
from its host, unless the recently detected stellar population 
in the LA (Casetti-Dinescu et al.\,2014; Zhang et al.\,2017) has
injected substantial amount of metals to increase the abundance in the gas.
Even then, the SMC-like abundance pattern in LA\,II today
is in line with the idea that LA\,II has been recently separated 
from the SMC (a few hundred Myr ago; see Pagel \& Tautvaivsiene 1998; 
Harris \& Zaritsky 2004). This scenario is also supported by the 
predominantly low $\alpha$-abundances observed in other regions of the LA 
(Fox et al.\,2018). We will further discuss these aspects in Sect.\,4.

\subsection{Molecular hydrogen abundance}

The molecular hydrogen absorption in the LA\,II is confined to 
comp.\,1, where also the C\,{\sc i} is observed. 
The total H$_2$ column density, as derived from summing over all
rotational levels $J$ (see Table 1), is log $N$(H$_2)=18.2\pm0.4$.
From the modeling of ATCA 21\,cm data follows 
that 80 percent of the observed neutral hydrogen column in LA\,II
of log $N=19.98$ resides in comp.\,1, so that log $N_1=19.88$.
The neutral gas column in the H$_2$/C\,{\sc i} bearing gas clump 
most likely is even smaller. This is evident from the lower $b$-value of 
the H$_2$/C\,{\sc i} absorption, which indicates the presence of 
a dense, spatially confined substructure in comp.\,1 that carries
only an (unknown) fraction of the total H\,{\sc i} column in
this component. The lower limit for the (local) fraction of hydrogen in 
molecular form in this sub-clump then is 
log $f_{\rm H_2}\equiv$\,log\,$[2N($H$_2)/(N$(H\,{\sc i}$)+2N($H$_2))]\geq -1.38$.
This logarithmic fraction is $0.65$ dex higher than the H$_2$ fraction
observed in the LMC filament of the MS towards Fairall\,9 (R13).
The interpretation of this difference requires a thorough modeling
of the H$_2$ formation-dissociation equilibrium, which will be discussed
in Sect.\,4.1 and in the Appendix.

\subsection{Dust depletion}

The significant underabundance of the elements Si, Fe, Al, and Ca in comp.\,1+2 and
comp.\,3 (see Table 2 and Sect.\,3.4.3) indicates the presence of interstellar dust grains
(e.g., Savage \& Sembach 1996).
Also the presence of H$_2$ in comp.\,1 hints at significant amounts of dust in LA\,II, because
hydrogen molecules can form efficiently only on the surfaces of dust grains 
(e.g, Spitzer 1978). For Fe, in addition, part
of the underabundance might be explained by nucleosynthetic
effects, namely the delayed production of Fe-peak elements from
core collapse and Type Ia supernovae (Tsujimoto et al. 1995).
Following the formalism described in R13, we define
for each element M the logarithmic depletion value
as log $\delta$(M)=\,[M/H$]_{\rm LA}-[$S/H$]_{\rm LA}$.
Using this definition, we derive the following the depletions values:
log $\delta$(Si$)=-0.5$ to $-0.6$ dex,
log $\delta$(Fe$)\geq -1.1$,
log $\delta$(Al$)=-1.2$, and
log $\delta$(Ca$)\geq -1.4$.
Due to the possible underabundance of Fe-peak elements in the gas,
only a lower limit for $\delta$(Fe) can be given.
If we take the Fe abundance in the SMC as reference, the Fe depletion
in LA\,II is log $\delta$(Fe$)=-1.0$.
For Ca, the upper limit for $\delta$(Ca) is a result of the 
unknown ionization correction for Ca\,{\sc ii}, which
strongly depends on the (unknown) gas density structure 
within comp.\,1+2 and thus cannot be constrained by our {\it Cloudy} model.

The (absolute) depletion values for Si and Al (and possibly also
Fe and Ca) are significantly larger than what we had found in the LMC 
filament of the MS towards Fairall 9, suggesting
that the dust-to-gas ratio in LA\,II is (at least locally) higher.
This is quite surprising, because the metallicity in LA\,II
is {\it lower} than in the LMC filament of the MS (R13).
These interesting findings will be further evaluated in Sect.\,5.2.


\section{Physical conditions in LA\,II}

\subsection{H$_2$ diagnostics}

Because of the slightly different values for $N$(H\,{\sc i}) and $N(J)$ for the 
H$_2$ compared to the earlier study of Wakker (2006) we here also update 
the subsequent determination of physical parameters in the H$_2$ gas,
such as density, temperature, thermal pressure.

For the interpretation of the relative H$_2$ abundance and the analysis of the H$_2$
rotational excitation, we follow exactly the same strategy as described in
R13. The most important equations are summarized in the Appendix.

If we balance the local UV flux in the LA\,II (based on the radiation
model presented in Fox et al.\,2014 assuming $d=20$ kpc) with a SMC-type H$_2$ grain-formation
rate (Tumlinson et al.\,2002) and further assume that H$_2$ line-self shielding applies, 
we obtain $\phi n_{\rm H}= 8$ cm$^{-3}$. Here, $\phi \leq 1$
describes the column-density fraction of the H\,{\sc i} in comp.\,1
that is physically related to the H$_2$ absorbing gas
(see Appendix for details). This density is roughly eight times higher
than the density in the H$_2$ bearing clump in the LMC-filament
of the MS (R13). Lacking any further information on the density distribution
in comp.\,1 we assume in the following $\phi =0.5$ as a realistic value (see R13),
so that we obtain $n_{\rm H}=16$ cm$^{-3}$.
From the 21\,cm ATCA observtions alone, WOP02 estimates (from geometrical 
considerations in the high-resolution interferometer data) a gas density of 
$n_{\rm H}=20$ cm$^{-3}$ in the densest cloud cores, thus in excellent
agreement with our value.

The thickness of the H$_2$ absorbing structure in the LA\,II then is given by 
$l_{\rm H_2}=\phi\,N$(H\,{\sc i}$)/n_{\rm H}\approx 0.8$ pc,
which is tiny compared to the overall size of LA\,II as seen
in 21\,cm emission (kpc-scale at a distance of $d=20$ kpc). 
Although the line of sight towards NGC\,3783 passes LA\,II in
a valley of the H\,{\sc i} column-density distribution (Fig.\,1), 
it obviously intersects a small, dense, dusty gas clump.


\begin{figure}[t!]
\epsscale{1.20}
\plotone{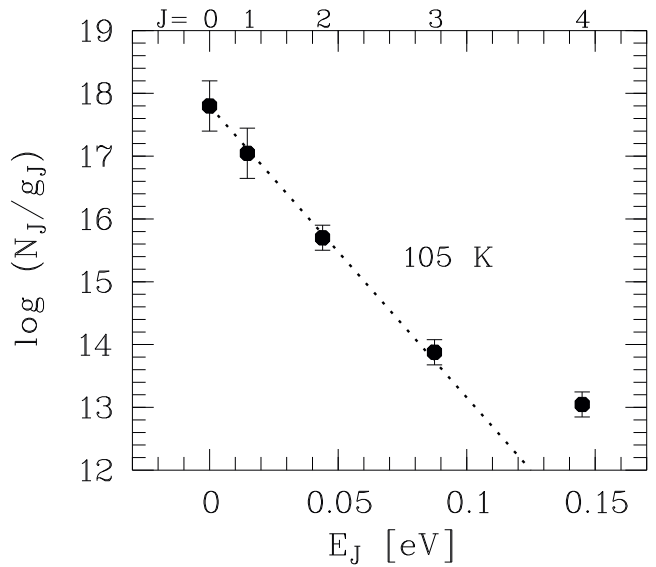}
\caption{
Rotational excitation of H$_2$ in the LA\,II. The H$_2$ column density for each rotational state,
divided by the state's statistical weight, is plotted against the state's excitation energy.
The data up to $J=3$ fit a single-component Boltzmann distribution with an excitation
temperature of $T_{\rm ex}=(105\pm 14)$ K.
}
\end{figure}


We now turn to the rotational excitation of the H$_2$,
which provides important information on the local gas temperature in the cloud core
(as in dense gas the lower rotational levels are excited by collisions).
In Fig.\,7 we show the H$_2$ column density for each rotational state, log $N(J)$,
divided by the state's statistical weight, $g_J$, against its
excitation energy, $E_J$. The distribution of the data points can be
fit for $J\leq 3$ with a single Boltzmann distribution with
$T_{\rm kin}\approx T_{\rm ex}=(105\pm 14)$ K, in Fig.\,7 shown as dashed line. Within the 
error range, this temperature is similar to values typically found in 
IVCs, in the MS towards Fairall\,9, and in the MCs (Richter 2000; Richter et al.\,2003a; 
Wakker 2006; R13; Tumlinson et al.\,2002), but is slightly higher than 
the canonical value of $T_{\rm kin}=77$ K that is characteristic for the 
MW disk (Savage et al.\,1977). But while in H$_2$-absorbing environments in the 
MW and in the MCs a two-component Boltzmann fit ususally is required to account for
photon-pumping processes in the higher $J$ levels ($J\geq 2$; e.g., Jura 1974), the 
H$_2$ rotational excitation in the LA\,II follows a different scheme with
just a single-component fit for $J=0-3$. This is not surprising, however, because the
local UV radiation field in the LA\,II is expected to be by a factor $\sim 35$ lower
compared to the MW disk (Fox et al.\,2014) due to the absence of a {\it significant}
(in terms of stellar luminosity density) local stellar population.
Therefore, photon-pumping
is not efficient in the LA\,II and the rotational excitation by collisions 
is the only revelant process that governs the distribution of $N(J)$ from the
ground states $J=0,1$ up to $J=3$.

With $R_{\rm LA}$ as SMC H$_2$ grain-formation rate density (see Appendix), the H$_2$ 
formation-time scale in the LA\,II towards NGC\,3783 is 
$(Rn_{\rm H})^{-1}\approx 640$ Myr. Depending on the origin and age of LA\,II,
the observed H$_2$ could be a relic of ancient molecule formation in the source
galaxy of the LA, or it could have formed in situ, with $f_{\rm H_2}$ adjusting to the
local physical conditions after the gas has been ripped off the
stellar body. These aspects will be further discussed in Sect.\,5.1.

With $T_{\rm kin}=105$ K and $n_{\rm H}=16$ cm$^{-3}$, the thermal pressure in the LA\,II
comes out to $P/k=n_{\rm H}T=1680$ cm$^{-3}$\,K (log $P/k=3.23$; see also Wakker 2006). 
This is more then than twice the pressure than what has been determined in the main body 
of the MS towards Fairall\,9 (R13). This significant pressure difference
most likely reflects the different locations of MS and LA in the halo with respect to the 
MW disk. With $d\leq 20$ kpc (Casetti-Dinescu et al.\,2014; Zhang et al.\,2017) the LA\,II is much closer
to the disk than the main body of the MS, which is assumed to be at $d\approx 50-100$ kpc.
From the models of Wolfire et al.\,(1995) it follows that a value of log $P/k=3.23$ 
for gas with an SMC chemical composition is above the thermal pressure component 
expected for a cold halo cloud located several kpc above/below the disk. 
If the ram pressure of the infalling high-velocity gas is included as an additional
pressure component, however, the derived value for $P/k$ in LA\,II is in line 
with the expectations.
For a characteristic infall velocity of $v_{\rm infall}=150$ km\,s$^{-1}$ (Heitsch \& Putman 2009), a value of log $P/k=3.23$
corresponds to a vertical height above the plane of $z\approx 10$ kpc (see Fig.\,5 in Wolfire et al.\,1995),
if a coronal gas temperature of $T=1-2\times 10^6$ K is assumed (Miller \& Bregman 2015). 
This $z$ height falls into the range expected from the observed stellar
distances in LA\,II (Casetti-Dinescu et al.\,2014, their Fig.\,3).
Therefore, it is plausible that the high gas density in the H$_2$ bearing clump 
and the resulting high value for $P/k$  reflect the proximity of LA\,II to 
the Milky Way disk and its compression by the  hydrodynamical interaction 
with the ambient coronal gas due to the high infall velocity.
Similar arguments have been used to explain the observed extremely high thermal gas
pressures in {\it outflowing} gas in the lower Galactic halo that most likely is associated 
with the Milky Way nuclear wind (Savage et al.\,2017). 
Unfortunately, a more precise estimate of the total gas pressure and the vertical position
of LA\,II above the plane, which is highly desired to further explore the possible 
interaction of LA\,II with the Milky way disk (McClure-Griffiths et al.\,2008), 
cannot be given at this point. This is because the ram pressure acting on the gas 
depends strongly on the space velocity of LA\,II, while only its radial velocity is 
securely constrained.


\begin{figure}[t!]
\epsscale{1.20}
\plotone{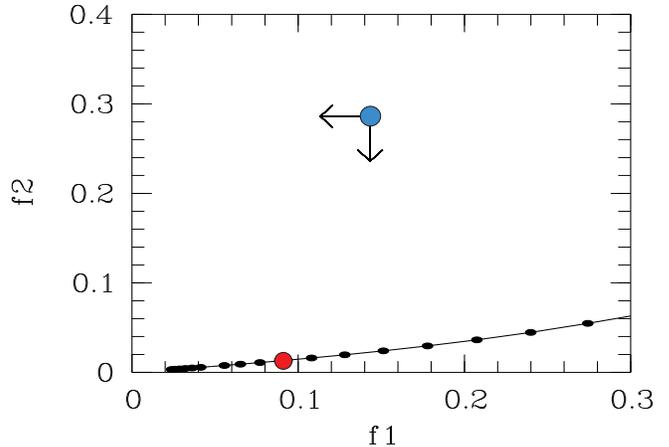}
\caption{
Limits on the relative abundances of the upper fine-structure levels of C\,{\sc i}
(parametrized by the quantities $f1$ and $f2$; see text) in
LA\,II towards NGC\,3783 (blue filled circle).
The solid black line shows the collisionally excitation model for $T=100$ K
from Jenkins \& Tripp (2001). The black dots on the line
indicate increasing logarithmic gas densities (from left to right)
in steps of 0.1 dex, while red solid point indicates the expected position
for log $n_{\rm H}=1.0$.
}
\end{figure}


\subsection{Fine-structure excitation of neutral and ionized carbon}

\subsubsection{Fine-structure excitation of C\,{\sc i}}

With an ionization potential of $11.2$ eV, neutral carbon can exist
in detectable amounts only in dense interstellar regions where a high
recombination rate can compensate for the photoionization of C\,{\sc i}.
The detection of C\,{\sc i} in comp.\,1 (Table 1) indicates the presence of
dense gas on its own, in line with above discussed gas properties derived from
the H$_2$ absorption. The measurement of the C\,{\sc i} in each of its three
fine-structure levels (C\,{\sc i}, C\,{\sc i}$^{\star}$, C\,{\sc i}$^{\star \star}$)
further allows a direct (independent) measurement of the thermal pressure in the gas,
as demosntrated in the local ISM (e.g., Jenkins \& Tripp 2001). Absorption
in C\,{\sc i}$^{\star}$ and  C\,{\sc i}$^{\star \star}$ is not detected
in LA\,II towards NGC\,3783, but upper limits have been derived (see Table 1).

Following the strategy discussed in Jenkins \& Tripp (2001) we define for
the relative abundance of the upper fine-structure levels the parameters
$f1=N($C\,{\sc i}$^{\star})/N($C\,{\sc i}$_{\rm tot}$) and
$f2=N($C\,{\sc i}$^{\star \star})/N($C\,{\sc i}$_{\rm tot}$). In Fig.\,8
we plot $f1$ vs. $f2$ and compare the position of this single data point
for LA\,II (blue filled circle) with
the collisional excitation model for $T=100$ K (solid line) from
Jenkins \& Tripp (2001; their Fig.\,5). The black dots on the line
indicate increasing logarithmic gas densities (from left to right)
in steps of 0.1 dex. The red solid point indicates the position
for log $n_{\rm H}=1.0$. As can be seen, no useful constraints on the
gas density/pressure in LA\,II can be obtained from the limits on
C\,{\sc i}$^{\star}$ and C\,{\sc i}$^{\star \star}$).
However, since this is one of the very few HVCs for which this experiment can be
carried out because of the very high S/N in the STIS data and the
detection of C\,{\sc i}, we show this diagnostic plot for sake of
completeness (see also Savage et al.\,2017 for another example)..


\begin{figure}[t!]
\epsscale{0.90}
\plotone{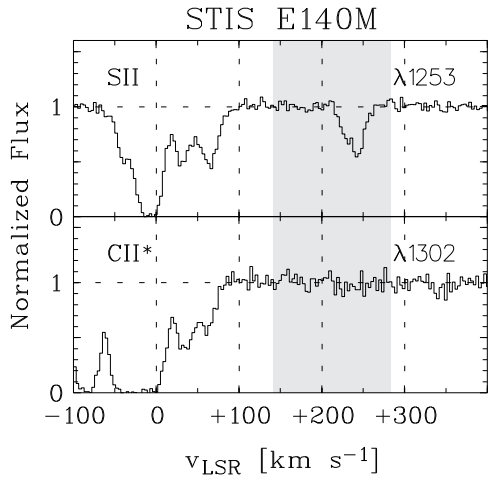}
\caption{
Velocity profile of C\,{\sc ii}$^{\star}$ $\lambda 1335.7$ absorption (lower panel) compared to
S\,{\sc ii} $\lambda 1253$. The gray-shaded area indicates the velocity range in which
absorption from LA\,II is expected. No significant C\,{\sc ii}$^{\star}$ absorption
is detected, however.
}
\end{figure}


\subsubsection{Fine-structure excitation of C\,{\sc ii}}

Because the relative population of the fine-structure levels of 
ionized carbon (C+) are governed by the balance between collisions 
with electrons and the radiative decay of the upper level into the 
ground state, the column-density ratio between C\,{\sc ii}$^{\star}$ and 
C\,{\sc ii} (or other singly ionized species) serves as measure for 
the electron density in different interstellar environments
(including HVCs and other CGM structures; Zech et al.\,2008; 
Jenkins et al.\,2005). In our previous study of the main body of the 
MS we have used the C\,{\sc ii}$^{\star}$/S\,{\sc ii} ratio 
to determine the electron density in the Stream towards
Fairall\,9 and C\,{\sc ii}$^{\star}$/H\,{\sc i} 
to estimate the local cooling rate in the gas (R13).

In Fig.\,9 we compare the velocity profile of C\,{\sc ii}$^{\star}$ $\lambda 1335$ 
and S\,{\sc ii} $\lambda 1253$ in our STIS spectrum of NGC\,3783.
No significant C\,{\sc ii}$^{\star}$ $\lambda 1335$ absorption
is detected at LA\,II velocities (gray-shaded) range, but a significant upper limit of 
log $N$(C\,{\sc ii}$^{\star})\leq 12.93$ can be measured from the data.
Following the strategy outlined in R13, we estimate from the measured
column-density ratio log [$N$(C\,{\sc ii}$^{\star})/N$(S\,{\sc ii}$)]\leq -1.47$
in comp.\,1 that log [$N$(C\,{\sc ii}$^{\star})/N$(C\,{\sc ii}$)]\leq -2.78$
(assuming a solar S/C ratio in the gas). For $T\approx 100$ K, (as 
measured from the H$_2$ rotational excitation), this is a very low value
and only reproducable in C\,{\sc ii} fine-struture population models
if the {\it average} gas density in comp.\,1 is by a factor of $\sim 5-10$
smaller than the peak value of $n_{\rm H}=16$ cm$^{-3}$ seen in 
the H$_2$-absorbing gas clump (see Keenan et al.\,1986; their Fig.\,2).
In any case, the upper limit for the electron density 
in comp.\,1 is $n_{\rm e}=0.01$ cm$^{-3}$ (Keenan et al.\,1986), 
a value that is characteristic for cold neutral gas at low metallicities
(e.g., Wolfire et al.\,1995) and comparable to what has 
been found in the MS towards Fairall\,9 ($n_{\rm e}\leq 0.05$ cm$^{-3}$; R13).

As mentioned above, the C\,{\sc ii}$^{\star}$/H\,{\sc i} ratio in LA\,II can
be used to estimate the local C$^{+}$ cooling rate, which
is governed predominantly by spontaneous deexcitations at these 
densities (Spitzer 1978). Following Lehner et al.\,(2004), the 
C$^{+}$ cooling rate per hydrogen atom can be calculated via the relation
$l_{\rm C}=2.89\times 10^{-20}\,N$(C\,{\sc ii}$^{\star})/N$(H\,{\sc i})
erg\,s$^{-1}$. For LA\,II, we derive as sightline average 
(integrating over all velocity components) an upper limit of 
log $l_{\rm C}\leq -26.62$. This value lies between the one
derived for the HVC Complex C ($\approx 0.15$ solar
abundance, log $l_{\rm C}\approx-27$; Lehner et al.\,2004; Richter et al.\,2001)
and the main body of the MS towards Fairall\,9 ($0.5$ solar abundance, 
log $l_{\rm C}\approx-26$; R13), which is not surprising given
the fact that LA\,II has a metallicity of $0.3$ solar.

\section{Discussion}

Our study confirms earlier results on the global metal abudance of 
LA\,II (Lu et al.\,1998; Fox et al.\,2018) and
provides new information on the dust properties and physical
conditions in the gas, further constraining of the origin of the LA
and its evolution.

\subsection{Physical properties and small-scale structure}

Taken together, the STIS, FUSE, and ATCA data indicate that the gas in LA\,II
in front of NGC\,3783 is multiphase, with several low-density components
and two dense, neutral cloud cores at $v_{\rm LSR}=245$ and $179$ km\,s$^{-1}$
(see Table 1). Comp.\,1, in particular, exhibits properties that are remarkable
for a Galactic halo cloud, with a very high gas density of $\sim 16$ cm$^{-3}$,
a low gas temperature of $T=105$ K, resulting in a high thermal pressure 
of $P/k\approx 1680$ K\,cm$^{-3}$, and a relatively high molecular gas fraction.
These results are in line with previous observations and consistent with
the detection of a stellar population in LA\,II (Casetti-Dinescu et al.\,2014),
indicating that LA\,II is situated at a distance of $d< 20$ kpc 
relatively close to the disk ($z\approx 10$ kpc), where the gas is compressed
and disrupted by ram pressure forces as a result of the interaction with the
ambient hot halo gas.
The physical properties and infall/disruption timescales mimic those seen in
hydrodynamical simulations of cold gas clouds moving into a hot plasma
at high speed (e.g., Heitsch \& Putman 2009; Joung et al.\,2012; 
Tepper-Garc{\'{\i}}a et al.\,2015; Armilotta et al.\,2016).

The H$_2$ bearing clump within LA\,II has a linear size of only $0.8$ pc (see
Sect.\,4.1), reflecting the clumpiness of the fragmenting neutral 
gas body of LA\,II. But also comp.\,3 at $+179$ km\,s$^{-1}$ 
appears to have a surprisingly high gas density of 
$n_{\rm H}\geq 0.3$ cm$^{-3}$ (Sect.\,3.4.3), which translates to an absorber 
thickness of just $l=N$(H\,{\sc i}$)/n_{\rm H}\leq 6$ pc, thus very small as well.
The formation of tiny, dense gas clumps due to the ablation of large
neutral gas complexes appears to be a phenomenon that is characteristic for the 
hydrodynamical interaction between infalling gas clouds and the ambient hot coronal gas,
as similar features have been observed also in other HVC complexes 
(see Richter et al.\,2003b, 2005b, 2009).

\subsection{Dust abundance}

One aspect that deserves further attention is the very high dust abundance
seen in LA\,II towards NGC\,3783 in both absorber groups. 
With the large depletion values of 
$\delta$(Si$)=-0.5$ to $-0.6$ dex, $\delta$(Fe$)\geq -1.1$ dex, $\delta$(Al$)=-1.2$ dex,
and $\delta$(Ca$)\geq -1.4$ dex
the dust content is exceptionally high for the low metallicity and the moderate 
neutral+molecular gas column density of log $N$(H$)\approx 20$. 
To put these results into a context, we compare the observed depletion
values with those found in the LMC and SMC.
Tchernyshyov et al.\,(2015) and Jenkins \& Wallerstein (2017) recently
have studied the depletion properties of various different elements in the
LMC and SMC using UV absorption-line data from HST/COS, HST/STIS, and FUSE.
In their studies, depletion values as high as in LA\,II are found in the 
LMC and SMC only in regions with total gas columns that are {\it substantially}
higher than in the LA\,II towards NGC\,3783.

To demonstrate this, we show in Fig.\,10 the Si depletion, $\delta$(Si), in
comp.\,1+2 and comp.\,3
as a function of the total neutral+molecular hydrogen column density, log $N$(H), for LMC, SMC, and LA\,II,
based on the results of Tchernyshyov et al.\,(2015) and this study. 
The Si depletion value in LMC and SMC increases strongly for decreasing hydrogen gas
column densities, as evident from the fits to the LMC and SMC data (solid red and blue lines
in Fig.\,6.) This overall trend is well known from observations in the Milky Way and
interstellar environments in other galaxies (Jenkins 2009), where the horizontal position
of the data points scales with the dust-to-gas ratio.
Obviously, the measured dust depletion in LA\,II towards NGC\,3783 in comp.\,1+2 and comp.\,3 
does not fit to the depletion trends in either LMC or SMC: it is apparently too
strong for the relatively low neutral+molecular hydrogen column density,
indicating a strongly enhanced local dust-to-gas ratio.
Fig.\,10, indicates that the level of depletion observed in LA\,II at 
log $N$(H$)=18.7$ and log $N$(H$)=20.0$ 
would correspond to a total neutral+molecular hydrogen column as high as 
log $N$(H$)\approx 21.2$ in the SMC (filled yellow box in Fig.\,10), with both 
environments having the same metallicity and overall abundance pattern.

There are two opposite ways to interpret this interesting deviation: either,
LA\,II has an unusually high fraction of heavy elements depleted into dust grains
when compared to LMC and SMC gas, or, for a given amount of dust, there is locally a
reduced neutral+molecular gas column density in LA\,II along the NGC\,3783 sightline 
(or a combination of both effects). We discuss both of these scenarios in the following.

{\it Enhanced dust growth.} 
Because most of the depletion of heavy elements into dust grains occurs due to dust
growth in the dense interstellar medium (e.g., Draine 2009; Jenkins et al.\,2009), it may 
well be possible that the large dust-to-gas ratio in LA\,II reflects an efficient dust 
growth {\it in situ}, possibly favoured by the high local gas density in comp.\,1 and 3 
and the absence of stellar and supernovae feedback that would lead to dust destruction.
It still appears remarkable (and not completely convincing to us) that even in dense CGM
environments the dust growth is so extremely efficient that a disk-like dust abundance
is achieved at column densities that are only a few percent of those observed in disks.

{\it Reduction of gas column density.} 
In this scenario, the neutral+molecular gas column density was substantially reduced 
(e.g., through hydrodynamical interaction with the ambient hot coronal gas)
after the gas was 
ripped off its mother galaxy, while the initial amount of dust stayed constant 
(or even was increased, see above scenario). If so, the stripped gas in LA\,II
would have ``stored'' the initial SMC/LMC-type relative dust abundance/depletion 
value (Fig.\,10), while the infalling cloud was breaking up and fragmenting into 
sub-clumps with a substantially reduced total neutral+molecular gas column density
(see Heitsch \& Putman 2009; Tepper-Garc{\'{\i}}a et al.\,2015).
Note that the dust-destruction timescale in the CGM far from
any supernova shocks and intense UV radiation fields is expeced to be very long, possibly
a few $10^8$ yr (Draine \& Salpeter 1979).

The 21\,cm map shown in Fig.\,1 does indeed demonstrate that the NGC\,3783 sightline
passes a {\it minimum} in the H\,{\sc i} column-density (CD) distribution of LA\,II,
lying between several clumps of higher CD (up to log $N$(H\,{\sc i}$)=20.6$; WOP02).
If the observed depletion level in LA \,II towards NGC\,3783 
reflects the initial dust conditions of the gas at the time when it was separated 
from the MCs, then the logarithmic neutral+molecular 
hydrogen column has dropped from $\sim 21.2$ to $\sim 20.0$ (comp.\,1+2+3; see Fig.\,10) 
along this 
sightline, as the cloud has moved into the inner Milky Way halo. This is a reduction
by more than 95 percent in neutral+molecular gas column. Given the long pathlength and 
the high radial velocity, such a cloud break-up and ablation of neutral gas column
is actually expected from hydrodynamical simulations 
that model the properties of neutral gas clouds of different initial masses
as they move at high speed through an ambient hot medium (Heitsch \& Putman 2009;
Joung et al.\,2012; Tepper-Garc{\'{\i}}a et al.\,2015).

While we here can only speculate about the origin of the high dust-to-gas ratio in LA\,II,
it represents an remarkable feature that deserves further attention in future studies.
Interestingly, previous analyses of the strongly depleted element Ca have demonstrated that 
dust commonly reaches out far beyond the stellar bodies of galaxies deep into circumgalactic 
and intergalactic space (e.g., Zhu \& M\'enard 2013; Richter et al.\,2011). In this broader context,
additional systematic studies of metal depletion in the Milky Way halo and in other 
circumgalactic environments would be very useful to better understand the large-scale 
circulation of dust in and around galaxies.

%

\begin{figure}[t!]
\epsscale{1.20}
\plotone{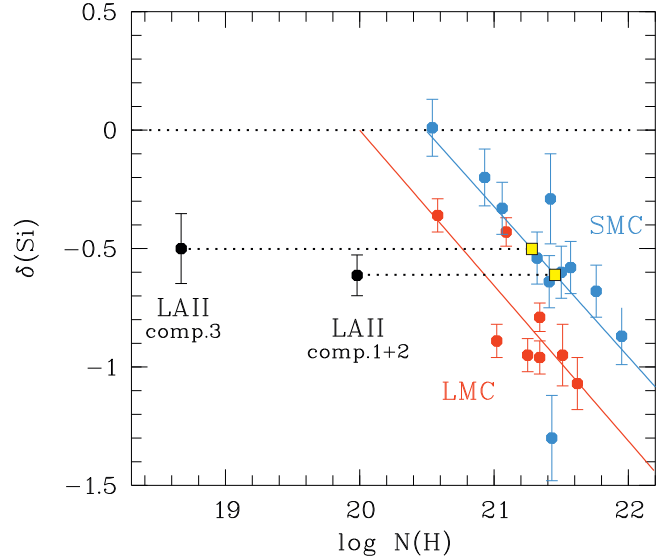}
\caption{
Si depletion values, $\delta$(Si), for LMC gas (filled red circles),
SMC gas (blue filled circles), and for LA\,II (filled black circles) in
comp.\,1+2 and comp.3 as function of the total atomic+molecular hydrogen
column density.
LMC and SMC data are adopted from Tchernyshyov et al.\,(2015). The
filled yellow box marks the hypothetical initial position of LA\,II for the
cloud-disruption model discussed in Sect.\,5.2.
}
\end{figure}


\subsection{Origin of LA\,II}

To further evaluate whether LA\,II originally stems from the SMC or LMC let us
first collect the main results from our absorption-line study and summarize the main 
conclusions from these findings:

\begin{enumerate}

\item LA\,II exhibits an $\alpha$ abundance and N/$\alpha$ ratio that closely
      matches that of the current-day SMC, possibly implying that the gas 
      recently was removed from 
      the SMC and pushed deep into the Milky Way halo.
      If so, the age of LA\,II can be at most a few hundred
      Myr. If we consider only the chemical abundance pattern as age indicator, 
      LA\,II could be older if the gas originates in the LMC. In that case,
      the observed abundance pattern in LA\,II would reflect that of the LMC 
      1-2 Gyr ago (Pagel \& Tautvaivsiene 1998; Harris \& Zaritsky 2004). That 
      the LMC had {\it exactly} the same $\alpha$/N abundance pattern as the SMC today 
      at the time when the gas was removed from the MCs is not unrealistic given 
      the star-formation history of the LMC (Pagel \& Tautvaivsiene 1998), but it
      would be a coincidence.

\item LA\,II exhibits a vey high thermal gas pressure of $P/k\approx 1680$ K\,cm$^{-3}$
      in the molecular gas-phase, 
      suggesting that the gas is situated in the {\it inner} halo of the Milky Way 
      $\approx 10$ kpc above the disk, where the gas is compressed 
      by ram-pressure forces. LA\,II and the MCs then would be separated by at least 30 kpc 
      in space, suggesting that it would have taken the gas more than 200 Myr to reach its current 
      position after it was separated from its host galaxy (assuming a relative speed of 
      $100$ km\,s$^{-1}$). This sets an minimum age of $\sim 200$ Myr to the LA\,II.
      Yet, a distance a small as $10$ kpc for the LA cannot be reproduced by the most recent
      hydrodynamical simulations of the Magellanic System (Pardy et al.\,2018) in a first infall 
      scenario, where the minimum distance of the LA is found to be $d\approx 24$ kpc.

\item The H$_2$ formation time-scale in the cold gas in LA\,II is $\sim 640$ Myr. If all 
      of the H$_2$ had formed in situ, this sets a lower limit for the age the LA\,II.
      However, given the fact that so much dust has survived the separation process,
      it is equally likely that also the H$_2$ (or at least part of it) had formed earlier 
      in the LA\,II source galaxy. If so, the H$_2$ was transported together with the neutral gas into
      the Milky Way halo, where the molecular gas fraction adjusted to the local conditions on much smaller
      time scales.

\item The very high dust-to-gas ratio in LA\,II suggests that the gas originally stems
      from a interstellar region with a neutral+molecular gas column as high as 
      log $N$(H)$=21.2$. Such high column densities are typical for the inner region in both LMC and 
      SMC (see, e.g., Br\"uns et al.\,2005; their Fig.\,5), but the average and peak 
      H\,{\sc i} column density is higher in the SMC than in the LMC. The high dust
      content in LA\,II strongly suggests that the gas was removed ``calmly'' from the MCs by
      ram-pressure forces and/or tidal forces, but not by ``violent'' processes, such
      energetic outflows from star-bursting regions that would have destroyed most
      of the dust (Feldmann 2015). 

\end{enumerate}

In summary, the observed chemical and physical properties do not exclude an LMC origin
of LA\,II, but they favor an SMC origin. In the latter case, the gas in LA\,II 
was removed from the SMC 
$200-500$ Myr ago during a close LMC-SMC encounter (Besla et al.\,2012;
Hammer et al.\,2015; Pardy et al.\,2018), followed
by a rapid decent into the Milky Way halo where the gas cloud was breaking up and
fragmented into smaller pieces due to the hydrodynamical interaction with the 
hot coronal gas.

If we combine the results from this study with those obtained from 
our recent HST/COS survey of the LA (Fox et al.\,2018) we find that
there is a substantial metallicity/age spread within the LA.
The lowest metallicity ($\sim 0.04-0.06$ solar) is found in LA\,III, followed by 
LA\,I, which has $\sim 0.12$ solar abundance, thus very similar to that in the 
main body of the MS (Fox et al.\,2013).
In contrast, the metallicity in LA\,II, as measured here, is with $\sim 0.3$ solar 
substantially higher than in the other LA complexes. Yet, it is still only half of that
measured in the LMC filament of the MS (R13). As discussed in Fox et al.\,(2018), 
the observed metallicity spread in the LA most likely indicates an episodic
removal of gas from the SMC, with LA\,III being the oldest part stripped 
from the SMC together with the main body of the MS $1-2$ Gyr ago, followed
by the later removal of LA\,I and LA\,II due to more recent encounters
of SMC and LMC. Our abundance measurements therefore set important new constraints for 
the numerical modeling of the the Magellanic System (see Pardy et al.\,2018). 
A detailed and systematic comparison between the observational results and 
the results from the numerical simulations would be of great importance to better 
understand, how the various components of the MS and LA came to their existence 
and what the observed gas properties tell us about the formation of the 
Magellanic System.

One other aspect that needs further attention in future LA studies is the role of
the LA's recently detected stellar population for the physical and chemical properties of the 
gas (Casetti-Dinescu et al.\,2014; Zhang et al.\,2017). While 
stars within the LA generally must lead to a local self-enrichment that may have the potential
to also increase the metal and dust abundance in the gas on larger spatial scales,
it remains to be evaluated in detail, how relevant this effect could be and whether 
the observed abundance variations in the LA could be related to such a local
stellar population.


\section{Summary and conclusions}

In this paper, we have presented new results on the chemical composition of 
the Leading Arm of the Magellanic Stream in the direction to the Seyfert galaxy
NGC\,3783, based on the combination of archival spectral data from
HST/STIS, FUSE, ATCA, GBT, and other instruments. 
Our main results can be summarized as follows:

\begin{enumerate}

\item Six individual velocity components belonging to LA\,II are identified in the 
      high-resolution STIS data at radial velocities between $+179$ and $+269$ km\,s$^{-1}$.
      There are two main absorber groups located at $+190$ and $+245$ km\,$^{-1}$ that 
      give rise to emission seen in the 21\,cm data. From the STIS data we construct an absorption-component
      model that we apply also to the lower-resolution FUSE and ATCA 21\,cm emission data.

\item From the modeling of the S\,{\sc ii}, O\,{\sc i}, and P\,{\sc ii} absorption lines in the STIS data and
      the analysis of the 21\,cm data from ATCA and GBT 
      we determine in both absorber groups a common $\alpha$ abundance in LA\,II of [$\alpha$/H$]=-0.53 \pm 0.08$,
      corresponding to $0.30\pm 0.05$ solar. This estimate includes an ionization
      correction based on {\it Cloudy} photoionization models of the gas in LA\,II.
      This value is similar to the present-day stellar $\alpha$ abundance 
      in the SMC. Our results confirm previous estimates of the LA\,II abundance based on 
      low-resolution  spectral data (Lu et al.\,1998), but increases substantially the 
      significance of this finding.

\item For the first time, we measure the abundance of nitrogen in the LA.
      The N abundance in LA\,II is very low, [N/H$]=-1.31 \pm 0.09$ ($0.05\pm 0.01$ solar) and
      [N/$\alpha]=-0.78$, again matching closely the present-day
      photospheric nitrogen abundance in the SMC. The elements Si, Fe, Al, and Ca are substantially 
      depleted into dust grains. We determine depletion values of $\delta$(Si$)=-0.5$ to $-0.6$ dex, 
      $\delta$(Fe$)\geq -1.1$ dex, $\delta$(Al$)=-1.2$ dex, and $\delta$(Ca$)\geq -1.4$ dex. 
      The substantial depletion values indicate that the
      local dust-to-gas ratio is very high for the only moderate H\,{\sc i} columns
      log $N$(H\,{\sc i}$)<20$ in both main absorber groups. We compare these values with 
      depletion patterns in LMC and SMC gas (Tchernyshyov et al.\,2015) and
      conclude that the gas in LA\,II presumably stems from a region that initially 
      had a high gas column density of log $N$(H$)\approx 21.2$.

\item The re-analysis of the H$_2$ absorption in the FUSE data yields column densities
      for the rotational levels, $N(J)$, that are very similar to those found in
      Wakker (2006). With a total H$_2$ column density of log $N$(H$_2)=18.22$ 
      the molecular fracion in the gas is $0.04$, which is (like the dust abudance)
      surprisingly high for the moderate H\,{\sc i} column density.
      We model the H$_2$ abundance in a formation-dissociation equilibrium including H$_2$
      self-shielding and a dissociating UV flux appropriate for the position of LA\,II in the
      Milky Way halo. From this we derive a high gas density of $n_{\rm H}=16$ cm$^{-3}$ in
      the cloud core of LA\,II. The rotational excitation of the H$_2$ yields a kinetic
      gas temperature of $T_{\rm kin}=(105\pm 14)$ K, implying that the thermal pressure in the molecular 
      gas is as high as $P/k\approx 1680$ K\,cm$^{-3}$.

\item The fact that the absolute and relative metal abundances in LA\,II exactly match those 
      observed in the SMC stellar population supports the scenario, in which LA\,II was stripped from the SMC during
      a recent LMC-SMC encounter $200-500$ Myr ago. However, an LMC origin as part
      of an earlier stripping process cannot be completely excluded from our data set. 
      The overall abundance spread in the LA presented in Fox et al.\,(2018), with LA\,II 
      being the most metal-rich region, points toward a periodic removal of gas from the SMC, 
      forming the chain of LA complexes seen today. The physical conditions in LA\,II and the high 
      dust abundance, 
      as derived in this study, indicate that the gas originally was of much higher column density
      before its was sinking into the Milky Way halo and being distrupted and dissolved due to 
      the interaction with the ambient hot halo gas. The high gas pressure in LA\,II
      suggests that LA\,II is only $\approx 10$ kpc away from the disk of the Milky Way
      in the lower halo or disk-halo interface, where ram-pressure forces add significantly to 
      the overall pressure budget on infalling neutral gas clouds.

\end{enumerate}

%

\acknowledgments

This work is based on observations with the NASA/ESA Hubble Space Telescope, obtained 
at the Space Telescope Science Institute (STScI), which is operated by the Association 
of Universities for Research in Astronomy, Inc., under NASA contract NAS 5-26555. Spectra 
were retrieved from the Barbara A. Mikulski Archive for Space Telescopes (MAST) at STScI. 
BPW was supported by NASA grants HST-GO-13448.01-A and HST-AR-14577.01-A. JCH and NL 
recognize support for this work from NASA and the NSF through grants HST-GO-14602 and 
AST-1517353, respectively. ED gratefully acknowledges the hospitality of the Center for 
Computational Astrophysics at the Flatiron Institute during the completion of this work.

%


\section*{REFERENCES}
\begin{footnotesize}

\noindent
Abgrall, H., \& Roueff, E. 1989, A\&A, 79, 313
\noindent
\\
Armilotta, L., Fraternali, F., \& Marinacci, F. 2016, MNRAS, 462, 4157
\noindent
\\
Asplund, M., Grevesse, N., Jacques Sauval, A., \& Scott, P. 2009, ARA\&A, 47, 481
\noindent
\\
Barger, K.A., Haffner, L.M., \& Bland-Hawthorn, J. 2013 ApJ, 771, 132
\noindent
\\
Barger, K.A., Madsen, G.J., Fox, A.J., et al.\,2017, ApJ, 851, 110
\noindent
\\
Besla, G., Kallivayalil, N., Hernquist, L., et al. 2010, ApJL, 721, L97
\noindent
\\
Besla, G., Kallivayalil, N., Hernquist, L., et al. 2012, MNRAS, 421, 2109
\noindent
\\
Boothroyd, A.I., Blagrave, K., Lockman, F.J., et al. 2011, A\&A, 536, A81
\noindent
\\
Brown, T., et al.\,2002, {\it HST} STIS Data Handbook, v.4, ed. B. Mobasher (Baltimore: STScI)
\noindent
\\
Br\"uns, C., Kerp, J., Stavely-Smith, L., et al. 2005, A\&A, 432, 45
\noindent
\\
Bustard, C., Pardy, S.A., D'Onghia, E., Zweibel, E.G., \& Gallagher, J.S.III 2018, ApJ, in press 
\noindent
\\
Casetti-Dinescu, Moni Bidin, C., K., Girard, T.M., et al.\,2014, ApJL, 784, L37
\noindent
\\
Dickey, J.M. \& Lockman, F.J. 1990, ARA\&A, 28, 215
\noindent
\\
D'Onghia, E., \& Fox, A.J. 2016, ARA\&A, 54, 363
\noindent
\\
Draine, B. \& Salpeter, E.E. 1979, ApJ, 231, 438
\noindent
\\
Draine, B., \& Bertoldi, F. 1996, ApJ, 468, 269
\noindent
\\
Draine, B. 2009, ASPC, 414, 453
\noindent
\\
Feldmann, R. 2015, MNRAS, 449, 3274
\noindent
\\
Ferland, G. J., Porter, R. L., van Hoof, P. A. M., et al. 2013, RMxAA, 49, 137
\noindent
\\
For, B.-Q., Staveley-Smith, L., McClure-Griffiths, N.M., Westmeier, T., \& Bekki, K. 
2016, MNRAS, 461, 892
\noindent
\\
Fox, A.J., Wakker, B.P., Smoker, J.V., et al. 2010, ApJ, 718, 1046
\noindent
\\
Fox, A.J., Richter, P., Wakker, B.P. et al. 2013, ApJ, 772, 110
\noindent
\\
Fox, A.J., Richter, P., Barger, K.A., et al.\,2014, ApJ, 787, 147
\noindent
\\
Fox, Anne, Richter, P., \& Fechner, C. 2014, A\&A 572, A102
\noindent
\\
Fox, A.J., Barger, K.A., Wakker, B.P., et al.\,2018, ApJ, 854, 142
\noindent
\\
Gabel, J.R., Crenshaw, D.M., Kraemer, S.B., et al.\,2003a, ApJ, 583, 178
\noindent
\\
Gabel, J.R., Crenshaw, D.M., Kraemer, S.B., et al.\,2003b, ApJ, 595, 120
\noindent
\\
Hammer, F., Yang, Y.B., Flores, H., Puech, M., \& Fouquet, S. 2015, ApJ, 813, 110
\noindent
\\
Harris, J., Zaritsky, D. 2004, AJ, 127, 1531
\noindent
\\
Heitsch, F. \& Putman, M. E. 2009, ApJ, 698, 1485
\noindent
\\
Herenz, P., Richter, P., Charlton, J.C., \& Masiero, J.R. 2013, A\&A, 550, A87
\noindent
\\
Howk, L.C., Wolfe, A.M. \& Prochaska, J.X. 2005, ApJL, 622, L81
\noindent
\\
Jenkins, E.B. \& Tripp, T.M. 2001, ApJS, 137, 297
\noindent
\\
Jenkins, E.B., Bowen, D.V., Tripp, T.M., \& Sembach, K.R. 2005, ApJ, 623, 767
\noindent
\\
Jenkins, E.B. 2009, ApJ, 700, 1299
\noindent
\\
Jenkins, E.B. \& Wallerstein, G. 2017, ApJ, 838, 85
\noindent
\\
Joung, M.R., Bryan, G.L., \& Putman, M.E. 2012, ApJ, 745, 148
\noindent
\\
Jura, M. 1974, ApJ, 191, 375
\noindent
\\
Keenan, F.P., Lennon, D.J., Johnson, C.T., \& Kingston, A.E. 1986, MNRAS, 220, 571
\noindent
\\
Lehner, N., Wakker, B.P., \& Savage, B.D. 2004, ApJ, 615, 767
\noindent
\\
Liang, C.J., \& Chen, H.-W. 2014, MNRAS, 445, 2061
\noindent
\\
Lu, L., Savage, B.D., \& Sembach, K.R. 1994, ApJ, 437, L119
\noindent
\\
Lu, L., Savage, B.D., Sembach, K.R., et al.\,1998, AJ, 115, 162
\noindent
\\
Mathewson, D. S., Cleary, M.N., \& Murray, J.D. 1974, ApJ, 190, 291
\noindent
\\
McClure-Griffiths, N.M., Staveley-Smith, L., Lockman, F.J. et al. 2008, ApJ, 673, L143
\noindent
\\
Miller, M.J., \& Bregman, J.N. 2015, ApJ, 800, 14
\noindent
\\
Morton, D. C. 2003, ApJS, 149, 205
\noindent
\\
Muzahid, S., Fonseca, G., Roberts, A., et al.\,2018, MNRAS, 476, 4965
\noindent
\\
Narayanan, A., Charlton, J.C., Masiero, J.R., \& Lynch, R. 2005, ApJ, 632, 92
\noindent
\\
Nidever, D. L., Majewski, S. R., \& Burton, W. B. 2008, ApJ, 679, 432
\noindent
\\
Nidever, D. L., Majewski, S. R., Burton, W. B., \& Nigra, L. 2010,
ApJ, 723, 1618
\noindent
\\
Pardy, S.A., D'Onghia, E., \& Fox. A.J. 2018, ApJ, 857, 101
\noindent
\\
Pagel, B. E. J., \& Tautvaisiene, G. 1998, MNRAS, 299, 535
\noindent
\\
Petitjean, P., Ledoux, C., Srianand, R. 2008, A\&A, 480, 349
\noindent
\\
Pettini, M., Zych, B. J., Steidel, C. C., \& Chaffee, F. H. 2008,
MNRAS, 385, 2011
\noindent
\\
Putman, M. E., Gibson, B. K., Staveley-Smith, L., et al.\,1998, Nature, 394, 752
\noindent
\\
Putman, M. E., Peek, J. E. G., \& Joung, M. R. 2012, ARA\&A, 50, 491
\noindent
\\
Richter, P. 2000, A\&A, 359, 1111
\noindent
\\
Richter, P., Savage, B. D., Wakker, B. P., Sembach, K. R., \& Kalberla, P. M. W.
2001, ApJ, 549, 281
\noindent
\\
Richter P., Wakker B. P., Savage B. D., \& Sembach K. R. 2003a, ApJ, 586, 230
\noindent
\\
Richter, P., Sembach, K.R., \& Howk, J.C. 2003b, A\&A, 405, 1013
\noindent
\\
Richter, P., Ledoux, C., Petitjean, P., \& Bergeron, J. 2005a, A\&A, 440, 819 
\noindent
\\
Richter, P., Westmeier, T., \& Br\"uns 2005b,  A\&A, 442, L49
\noindent
\\
Richter, P. 2006, Rev.\,Mod.\,Astron., 19, 31
\noindent
\\
Richter, P., Charlton, J.C., Fangano, A.P.M., Ben Bekhti, N.,
\& Masiero, J.R. 2009, ApJ, 695, 1631
\noindent
\\
Richter, P., Krause, F., Fechner, C., Charlton, J.C., \& Murphy, M.T. 2011, A\&A, 528, A12
\noindent
\\
Richter, P., Fox, A.J., Wakker, B.P., et al.\,2013, ApJ, 772, 111 
\noindent
\\
Richter, P., Wakker, B.P., Fechener, C., Herenz, P., Tepper-Garc\'ia, T., \& Fox, A.J. 
2016, A\&A, 590, A68
\noindent
\\
Richter, P., Nuza, S. E., Fox, A. J., et al.\,2017, A\&A, 607, A48
\noindent
\\
Richter, P. 2017, ASSL, 430, 15
\noindent
\\
Savage, B.D., Drake, J.F., Budich, W., \& Bohlin, R.C. 1977, ApJ,
216, 291
\noindent
\\
Savage, B.D. \& Sembach, K.R. 1991, ApJ, 379, 245
\noindent
\\
Savage, B.D. \& Sembach, K.R. 1996, ARA\&A, 34, 279
\noindent
\\
Savage, B.D., Kim, T.-S., Fox, A.J., et al.\,2017, ApJ, 232, 25
\noindent
\\
Sembach, K.R., Howk, J.C., Savage, B.D., \& Shull, J.M. 2001,
AJ, 121, 992
\noindent
\\
Spitzer, L. 1978, {\it Physical processes in the interstellar medium}, 
(New York Wiley-Interscience)
\noindent
\\
Stocke, J.T. Keeney, B.A., Danforth, C.W., et al.\,2014, ApJ, 791, 128
\noindent
\\
Tchernyshyov, K., Meixner, M., Seale, J., et al. 2015, ApJ, 811, 78
\noindent
\\
Tepper-Garc{\'{\i}}a, T., Bland-Hawthorn, J., \& Sutherland, R.S. 2015, ApJ, 813, 94
\noindent
\\
Tsujimoto, T., Nomoto, K., Yoshii, Y., et al. 1995, MNRAS,
277, 945
\noindent
\\
Tumlinson, J., Shull, J. M., Rachford, B. L., et al. 2002, ApJ, 566, 857
\noindent
\\
Venzmer, M.S., Kerp, J., \& Kalberla, P.M.W. 2012, A\&A, 547, 12
\noindent
\\
Wakker, B. P., Oosterloo, T. A., \&  Putman, M. E. 2002, AJ, 123, 1953
\noindent
\\
Wakker, B.P. 2006, ApJS, 163, 282
\noindent
\\
Wannier, P., \& Wrixon, G. T. 1972, ApJ, 173, L119
\noindent
\\
Werk, J.K., Prochaska, J.X., Thom, C., et al.\, ApJS, 2014, 17
\noindent
\\
West, K.A., Pettini, M, Penston, M.V., Blades, J.C., \& Morton, D.C. 1985, MNRAS, 215, 481
\noindent
\\
Wolfire, M.G., McKee, C F., Hollenbach, D., \& Tielens, A.G.G.M.
1995, ApJ, 453, 673
\noindent
\\
Zech, W.F., Lehner, N., Howk, J.C., Dixon, W.V.D., \& Brown, T.M. 2008, ApJ, 679, 460
\noindent
\\
Zhang, L., Moni Bidin, C., Casetti-Dinescu, D.I., et al.\,2017, ApJ, 835, 285
\noindent
\\
Zhu, G. \& M\'enard, B. 2013, ApJ, 773, 16 
\end{footnotesize}

%

\newpage

\appendix

\section{H$_2$ analysis}


\begin{figure}[h!]
\epsscale{0.70}
\plotone{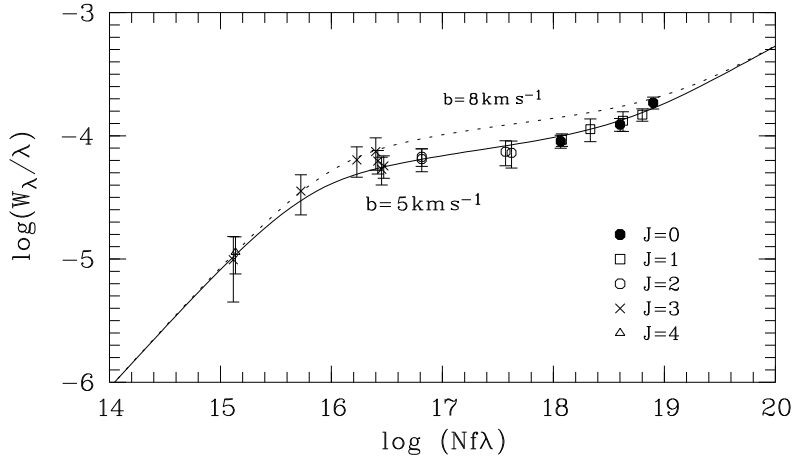}
\caption{
Curves of growth for H$_2$ absorption in comp.\,1. The absorption lines
arising from the rotational levels $J=0-2$ fit best on a curve of
growth with $b=5.0$ km\,s$^{-1}$. For $J=3$, the scatter of the data points
also allow higher $b$-values in the range  $b=5-8$ km\,s$^{-1}$.
}
\end{figure}


\subsection{Curve-of-growth}

To independently constrain the $b$-value for
the H$_2$ absorption in the LA\,II we have measured the equivalent widths
for a sub-set of unblended H$_2$ lines that have particularly high S/N
in the FUSE data and constructed curves of growth (CoG; Fig.\,A.1).
As can be seen, the CoG method further supports a $b$ value of
$b=5.0$ km\,s$^{-1}$ for the H$_2$ , at least for the lower rotational
levels $J=0-2$. For $J=3$, the scatter in the data points
is larger than for $J\le 2$, so that any value in the range $b=5.0 - 8.0$ km\,s$^{-1}$ 
is possible. If $b>5.0$ km\,s$^{-1}$ for $J=3$, this
would indicate that the rotationally excited
H$_2$ gas is spatially more extended than the ground-state H$_2$,
which is plausible within the scenario of a core-envelope structure
of comp.\,1. For the determination of the $J=3$ column density,
the choice of $b$ within the range $b=5.0-8.0$ km\,s$^{-1}$ is
not important, as $N(3)$ is contrained mostly by the very weak lines
on the linear part of the CoG.

\subsection{H$_2$ diagnostics}

For the analysis of the H$_2$ absorption we follow the same
strategy as adopted for the H$_2$ absorption in the main body
of the MS towards Fairall\,9 (R13).
The relative abundance of H$_2$ compared to H\,{\sc i}
is regulated by the formation-dissociation equilibrium
of molecular hydrogen, which can be expressed as

\begin{equation}
\frac{N({\rm H\,I})}{N({\rm H}_2)} =
\frac{\langle k \rangle \,\beta}{R\,n({\rm H\,I})\,\phi}.
\end{equation}

\noindent
In this equation, $\langle k \rangle \approx  0.11$ is the
probability that the molecule is dissociated after photo absorption.
The parameter $\phi \leq 1$
in equation (A1)
describes the column-density fraction of the H\,{\sc i}
that is physically related to the H$_2$ absorbing gas,
i.e., the fraction of neutral gas atoms that can
be transformed into H$_2$ molecules (see Richter et al.\,2003a
for details).

The parameter $\beta$ is the H$_2$ photo-absorption rate per second
within the H$_2$-bearing cloud, while $R$ represents the H$_2$ formation
rate on dust grains in units cm$^{3}$\,s$^{-1}$.
Estiming $\beta$ for the LA is non-trivial, because one needs to
estimate the dissociating photon flux at the position of the LA in the halo
and, additionally, consider the effect of H$_2$ line-self shielding {\it within}
the cloud. From the radiation-field model by Fox et al.\,(2014) it follows that
the FUV photon flux at the position of the LA in the halo at distance 20 kpc
is reduced by a factor of $35$. For the H$_2$ photo-absorption rate in the Milky Way disk, we
assume the canonical value of $\beta_{\rm 0,MW}=5.0\times 10^{-10}$ s$^{-1}$
(e.g., Spitzer 1978), which implies that $\beta_{\rm 0,LA}=1.4\times 10^{-11}$ s$^{-1}$
at the edge of the clump in the LA towards NGC\,3783. The H$_2$ line-self shielding
further reduces this absorption rate in the interior of the clump. Following the
parametrization scheme presented by Draine \& Bertoldi (1996), the absorption
rate in the cloud can be expressed as $\beta=S\,\beta_0$,
where $S=(N_{\rm H_2}/10^{14}$cm$^{-2})^{-0.75}<1$. From this we finally
obtain $\beta=1.0\times 10^{-14}$ s$^{-1}$ in the self-shielded cloud core.

The formation of H$_2$ predominantly takes place on the surface of dust grains,
as pure gas-phase reactions of hydrogen atoms and ions are far
less efficent than the formation on dust (e.g., Spitzer 1978).
For the required H$_2$ grain formation rate in the LA, we adopt
the value determined for the SMC (see Tumlinson et al.\,2002),
$R_{\rm LA} = 3 \times 10^{-18}$ cm$^{3}$\,s$^{-1}$, which
is appropriate given the SMC-like chemical abundance pattern in LA\,II
and the observed dust-depletion pattern (Sect.\,3.6).
From equation (A1) we then derive $\phi n_{\rm H}\approx 8$ cm$^{-3}$.

\section{Comparison between HST/STIS and HST/COS data}

The analysis of the COS G130M data of NGC\,3783 presented in 
Fox et al.\,(2018) yields a slightly lower S\,{\sc ii} column
density than our analysis of the STIS E140M data
(COS: log $N$(S\,{\sc ii}$)=14.41\pm 0.12$; 
STIS: log $N$(S\,{\sc ii}$)=14.57\pm 0.03$),
although both 
studies use the same S\,{\sc ii} line at $1253.8$ \AA.
In Fig.\,B1 we compare the STIS E\,140M data with the COS G130M data
in the spectral region around the S\,{\sc ii} $\lambda 1253.8$ line
to explore the origin for this discrepancy. As can be seen, the 
COS data are slightly more noisy than the STIS data. There also are
some continuum undulations present and less absorption is seen 
in the line centers in the COS data, particularly in the region of the LA\,II
absorption near $1254.8$ \AA. The observed discrepancies between
both data sets can be explained by the lower spectral resolution
of the COS data ($R\approx 15,000$ for COS vs. $R\approx 45,000$ for
STIS) in combination with COS fixed-pattern features that are
present in many COS data sets (e.g., Richter et al.\,2017).


\begin{figure}[t!]
\epsscale{0.50}
\plotone{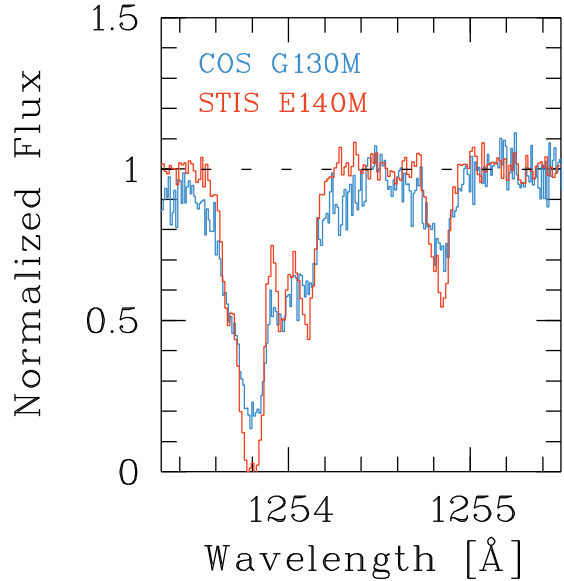}
\caption{
Comparison between STIS E\,140M data (red) and COS G130M data (blue) of 
NGC\,3783 in the wavelength range between $1253.3$ and 
$1255.5$ \AA, covering the S\,{\sc ii} $\lambda 1253$ line.
}
\end{figure}


%

\end{document}